\documentclass[10pt,journal,compsoc]{IEEEtran}
\usepackage{times}
\usepackage{epsfig}
\usepackage{graphicx}
\usepackage{amsmath}
\usepackage{amssymb}
\usepackage[ruled]{algorithm2e}
\usepackage{algpseudocode}
\usepackage{caption}
\usepackage{subfigure}
\usepackage{booktabs}
\usepackage{tabularx}
\usepackage{xcolor}
\usepackage{multirow}
\usepackage{subfig}
\usepackage{cleveref}

\newcolumntype{Y}{>{\centering\arraybackslash}X}

\begin{document}
%

\title{SecureSense: Defending Adversarial Attack for Secure Device-Free Human Activity Recognition}


\author{Jianfei~Yang,
 	Han~Zou,
 	and~Lihua~Xie,~\IEEEmembership{Fellow,~IEEE}
 \IEEEcompsocitemizethanks{
	\IEEEcompsocthanksitem J. Yang and L. Xie are with the School of Electrical and Electronics Engineering, Nanyang Technological University, Singapore (e-mail: yang0478@ntu.edu.sg; elhxie@ntu.edu.sg).
 
 	\IEEEcompsocthanksitem H. Zou is with the Department of Electrical Engineering and Computer Sciences, University of California, Berkeley, USA (e-mail: hanzou@berkeley.edu).
}
}

\markboth{IEEE Transactions on Mobile Computing}%
{Shell \MakeLowercase{\textit{et al.}}: Bare Demo of IEEEtran.cls for IEEE Journals}

\maketitle

\begin{abstract}
   Deep neural networks have empowered accurate device-free human activity recognition, which has wide applications. Deep models can extract robust features from various sensors and generalize well even in challenging situations such as data-insufficient cases. However, these systems could be vulnerable to input perturbations, i.e. adversarial attacks. We empirically demonstrate that both black-box Gaussian attacks and modern adversarial white-box attacks can render their accuracies to plummet. In this paper, we firstly point out that such phenomenon can bring severe safety hazards to device-free sensing systems, and then propose a novel learning framework, SecureSense, to defend common attacks. SecureSense aims to achieve consistent predictions regardless of whether there exists an attack on its input or not, alleviating the negative effect of distribution perturbation caused by adversarial attacks. Extensive experiments demonstrate that our proposed method can significantly enhance the model robustness of existing deep models, overcoming possible attacks. The results validate that our method works well on wireless human activity recognition and person identification systems. To the best of our knowledge, this is the first work to investigate adversarial attacks and further develop a novel defense framework for wireless human activity recognition in mobile computing research.
\end{abstract}



\section{Introduction}
\IEEEPARstart{H}{uman} Activity Recognition (HAR) has aroused much attention in mobile computing and smart sensing research. Classic HAR technology is empowered by various Internet of Things (IoT) sensors, such as Inertial Measurement Unit (IMU) that has been equipped in most smartphones for gait counting \cite{chen2017robust}. However, it is not always convenient to employ device-based HAR, especially at smart homes, which gives birth to device-free HAR that leverages radar \cite{li2019survey}, WiFi \cite{yang2018device} and ultrasound \cite{jiang2018towards} for sensing. With the recent booming development of deep learning, fine-grained activities patterns can be extracted and recognized from device-free sensor data. For example, relying on state-of-the-art convolutional and recurrent neural networks, human activity and gesture recognition techniques are empowered using off-the-shelf WiFi devices \cite{zhang2021gaitsense,yang2019learning,yang2018carefi,yang2018fine,deng2022gaitfi}.

\begin{figure*}[t]
	\centering
	\includegraphics[width=0.8\linewidth]{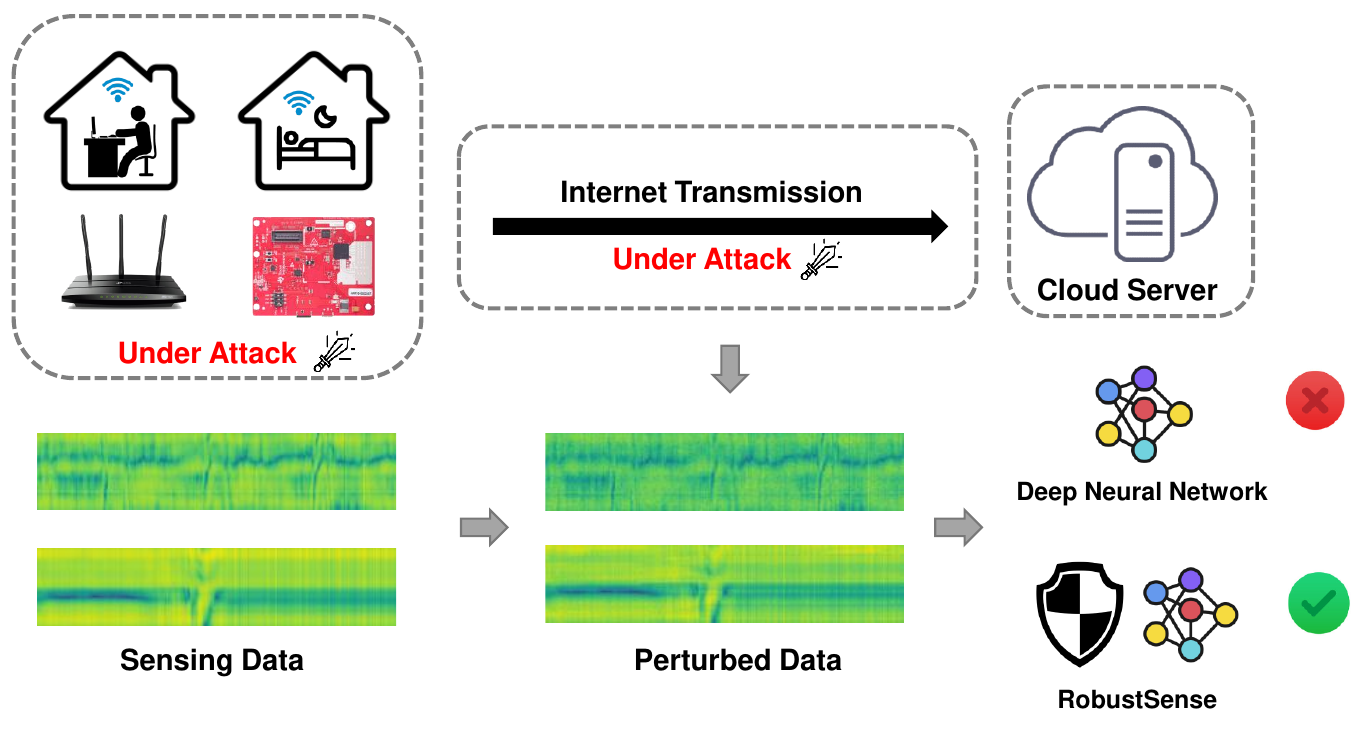}
	\caption{The illustration of the proposed SecureSense framework. It successfully defends Gaussian noise attack and adversarial attack. Furthermore, defending attacks boosts model robustness and even achieves better recognition performance.}
	\label{fig:intuition}
\end{figure*}

Though deep models show superior performance for device-free HAR, they require powerful computational resources, i.e. Graph Processing Unit (GPU) or Tensor Processing Unit (TPU). This situation requires users to transmit sensing data to a cloud server for deep model inference, and then results are returned to user interface via Internet. During such transmissions, sensing data could suffer from various potential adversarial attacks. Note that deep neural networks are very sensitive to these data perturbations \cite{szegedy2013intriguing}. Furthermore, even based on edge computing without transmission to cloud server, such adversarial attacks can directly happen on IoT devices due to insecure IoT operating systems and vulnerable firewalls of edge devices. These attacks can fool the deep models to obtain wrong or even hacker-defined results, leading to intractable situations in the real world \cite{madry2017towards}.

Adversarial attacks have been explored in computer vision research \cite{akhtar2018threat}, where severe consequences may be caused by these attacks for face recognition or autonomous driving systems. Random noise added to raw data is the simplest attack approach, leading to degraded performance of deep models but this can be alleviated by smoothing classifier or data augmentation \cite{zhao2021expressive}. However, more hazardous blind spots are adversarial examples generated by attack algorithms, which confuse model prediction or even induce deep models to generate specific wrong results, which leads to huge threats to public vision systems \cite{goodfellow2014explaining}. Researchers have developed dedicated approaches to defend different kinds of attacks for secure visual recognition algorithms \cite{yuan2019adversarial}.

{\color{black}
Despite great progresses in computer vision \cite{goodfellow2014explaining} and natural language processing \cite{zhang2020adversarial}, there is a lack of research on data attack and defense for device-free HAR systems. It is noted that HAR systems also highly demand safety and security. For instance, in smart homes, spiteful tamper of sensing data can deceive the access control system, leading to a severe security problem. In smart buildings,  illegal manipulation of massive occupancy data hinders the occupancy estimation system, increasing energy assumption or even triggering emergency equipment such as fire-fighting systems. Normally, to defend adversarial attacks, data augmentation and adversarial training help deep models generalize well \cite{madry2017towards}. However, for deep HAR models under attack, we find that these two methods only result in a marginal improvement. Furthermore, instability of training procedure and hyper-parameter tuning bring enormous difficulties to a realistic secure deep recognition model for device-free HAR systems. 

In this paper, we investigate the scenario of devide-free HAR when an attacker manages to tamper sensing data by adding artificial noise during communication so that the HAR model can make wrong prediction and then interfere with downstream applications, such as the fire-fighting system mentioned above. To begin with, we study how IoT-enabled device-free HAR systems perform under different adversarial attacks. It is found that adversarial attacks have serious impact on the performances of deep HAR models. Attacks to a single sensor dimension have brought about dramatic performance dropping, and mixed attacks lead to worst performance. To enhance the attack robustness of deep models, we develop a novel method, SecureSense, which shows consistent performance for both normal and adversarial data samples with a stable training procedure. As shown in Figure \ref{fig:intuition}, the SecureSense aims to learn consistent predictions for the raw input and series of simulated attack data, achieving state-of-the-art defense performance. For HAR situations, we propose a more challenging attack that integrates black-box and white-box attack, termed as bimodal attack. In the evaluation, we prototype our method on wireless human activity recognition and human identification systems. Extensive experiments demonstrate that the SecureSense can defend these attacks with unseen hyper-parameters.

The contributions of this work are summarized as follows:
\begin{itemize}
	\item We firstly investigate how adversarial attacks affect the safety and security of existing deep models running at HAR systems, demonstrating their vulnerabilities by existing adversarial and the proposed bimodal attack. To the best of our knowledge, this work is the first that studies the adversarial examples and defense for device-free HAR systems.
	\item We propose a novel adversarial defense approach, namely SecureSense, which leverages prediction consistency between normal and adversarial examples as a regularization for better model robustness.
	\item We conduct extensive experiments on wireless human activity recognition and authentication systems. The SecureSense offers significant improvement of recognition accuracy with a stable training procedure, avoiding cumbersome hyper-parameter tuning process for adversarial training.
\end{itemize}
}

The rest of this paper is organized by five sections. Section \ref{sec:attack} introduces existing IoT-enabled device-free sensing systems and conduct a preliminary research on adversarial attacks. Then we introduce our defending method in Section \ref{sec:defense}, which is followed by the experiments in Section \ref{sec:experiment}. The related works are reviewed in Section \ref{sec:related-work} and we conclude the paper in Section \ref{sec:conclusion}.




\section{Security Issue in Device-Free Sensing}\label{sec:attack}
\subsection{IoT-enabled Device-Free Sensing System}
Recent IoT-enabled sensors and deep learning models have empowered accurate device-free human activity recognition. The device-free HAR sensors are mainly composed of light, ultrasound, radar and radio frequency. These HAR systems usually collect and transmit data to an embedded IoT board or a cloud server for further computations, which is vulnerable to attackers. Attackers can hack in communication system or IoT operating system to tamper sensing data. By adding some artificial noises, sensing data becomes ``adversarial samples'' that induces deep models to make wrong or even attacker-specified predictions, leading to misleading data in downstreaming applications of HAR. For example, the goal of the attackers could be to disturb the occupancy estimation and fire alarm system. The wrong human data can lead to more energy consumption and trigger fire-fighting systems. This paper aims to study how to increase the model security by enhancing the prediction robustness against these adversarial samples.

Among prevailing device-free HAR systems, WiFi-based sensing leverages existing WiFi infrastructures for smart sensing, which is cost-effective. Furthermore, it extracts fine-grained Channel State Information (CSI) from Multiple Input Multiple Output (MIMO) WiFi communication links \cite{xie2015precise,yang2022deep,yang2021deep}, and enables accurate activity or gesture recognition applications \cite{yang2018device,zou2018deepsense,wang2021multimodal}. Here, we set WiFi-based device-free sensing system as an example, and study how adversarial attacks affect their deep HAR models. In wireless sensing, the WiFi signals are transmitted from a transmitter to a receiver, and the CSI data reflects the propagation states of communication links such as reflection and diffraction phenomena of WiFi signals. When human activities are performed within the sensing range, specific human motions can lead to similar patterns of CSI curves. In this sense, different human activities or gestures have different CSI patterns, which can be identified by deep neural networks \cite{yang2019learning}. The CSI data is sampled from Channel Impulse Response (CIR) $h(\tau)$ in Orthogonal Frequency Division Multiplexing (OFDM) at the physical layer, and can be represented by
\begin{equation}
	H_i=||H_i||e^{j \angle H_i},
\end{equation}
where $||H_i||$ and $\angle H_i$ are the amplitude and phase of $i$-th subcarrier. For each timestamp, we have a CSI sample that consists of 56 subcarriers at 40MHz bandwidth and 114 subcarriers at 40MHz bandwidth for one pair of antennas. Running a WiFi-based HAR system with 3 pairs of antennas at 80MHz bandwidth, we can collect activity data samples $x$ with a size of $3\times 114\times T$ where $T$ is the number of packets representing the time duration of an activity. $T$ normally stands for a few seconds duration with a sampling rate of 10-100Hz, and falls into $[50,500]$. Since the amplitude of CSI has been sufficient for HAR, only amplitude modality is utilized in this paper.

To recognize the CSI data, a deep neural network $h(x)$ parameterized by $\theta_h$ is constructed for activity recognition by an aggregation of various layers including convolutional layers, fully-connected layers, pooling layers, etc. The output of $h(x)$ is the predicted activity category $\hat{y}$ in the label space $Y={1,2,3,\dots,K}$ where $K$ is the number of recognizable human activities. After training the model $f(x)$ using massive labeled data $\{x_i, y_i\}^N_{i=1}$ where $N$ is the number of data samples, we obtain an accurate deep HAR model by optimizing the activity classification loss w.r.t. the model parameters $\theta_h$. 

\subsection{Adversarial Attacks for Device-Free HAR Systems}\label{subsec:attack}
The existence of adversarial examples can be caused by linear behavior of deep models in high-dimensional space, though they have strong capacity of fitting non-linear data \cite{goodfellow2014explaining}. For a CSI data sample $x \in \mathbb{R}^M$ and a perturbation $\zeta \in \mathbb{R}^M$, an adversarial example $\tilde{x}$ is constructed by
\begin{equation}\label{eq:adversarial-example}
	\tilde{x} = x + \zeta.
\end{equation}
Feeding such an adversarial example into a linear layer $\phi(x)=w^\intercal x+b$, we obtain
\begin{equation}
	w^\intercal \tilde{x} = w^\intercal x + w^\intercal \zeta.
\end{equation}
Obviously, the activation of $\phi(x)$ is increased by $w^\intercal \zeta$. Though each element in $\zeta$ might be small, $w^\intercal \zeta$ can sum up these perturbations and make a tremendous difference to the activation, especially for high dimensional space with a large $M$. In wireless sensing, the data dimension ranges in $3\times 114\times T$ where $T\in[50,500]$. Ultrasound and radar HAR systems also come with high-dimensional data spaces \cite{li2019survey}. Therefore, it is noticeable that such high dimensional HAR data could be vulnerable to adversarial examples.

{\color{black}
\subsubsection{Bimodal Attack}
Adversarial examples can be generated by either black-box or white-box strategy. Black-box attack refers to the attacker who cannot access to the machine learning model, while white-box attack is conducted by leveraging the parameters of the model. Here we propose a new attack scheme, namely bimodal attack, which integrates both black-box and white-box strategies. Bimodal attack is based on perturbations on sensing data via Gaussian noise and adversarial training, and it shows more harmful for HAR tasks. 

To develop bimodal attack, we firstly introduce Gaussian Noise (GN) attack. Directly adding random noise to data is a simple approach of generating adversarial examples. Let the perturbation $\zeta$ follow the Gaussian distribution with mean $\mu$ and standard deviation $\sigma$, i.e. $\zeta \sim \mathcal{N}(\mu,\sigma^2)$. The perturbed CSI data $\tilde{x}$ can effectively degrade the deep HAR models. Since the model $h(x)$ is not required to generate adversarial examples, GN attack is regarded as a black-box attack strategy.

If the attacker has access to the HAR model or the training data, then the white-box attack strategy is more effective. Here we introduce the famous Fast Gradient Sign Method (FGSM) \cite{goodfellow2014explaining}, a simple yet effective method to deceive deep models. Let $L(x,y;\theta_h)$ be the classification loss to train the HAR network $f(x)$. An optimal max-norm constrained perturbation is obtained by
\begin{equation}
	\zeta = \epsilon \text{sign} \big(\nabla_x L(x,y;\theta_h) \big),
\end{equation}
where $\nabla$ is the partial derivative of $L(x,y;\theta_h)$ to $x$. The gradient is easily obtained by backpropagation when training the neural network $h(x)$. In computer vision, it is observed that a small $\epsilon$ can easily degrade the performance of deep HAR models which could be even lower than the accuracy of random guess. In our evaluation, similar situation happens to wireless HAR systems.

Since GN and FGSM attacks affect deep HAR models from two different perspectives, combining two attacks should have a more harsh impact. Relying on this idea, we propose the bimodal attack that aggregates the FGSM and GN attacks in a cascade fashion. In bimodal attack, the FGSM is firstly applied and then GN will be superposed to each entry value of the input matrix, as defined by:
\begin{equation}
	\zeta = \Psi_{\mathcal{N}} + \epsilon \text{sign} \big(\nabla_x L(x,y;\theta_h) \big),
\end{equation}
where $\Psi_{\mathcal{N}}$ denotes a random variable that follows the Gaussian distribution. It is discovered that our bimodal attack is more difficult to defend as due to the mixture of white-box and black-box attacks.

\subsubsection{Antenna Level Attack}
The HAR sensing data has two dimensions -- sensor dimension and time duration. The time duration is flexible for this kind of applications via sliding window, while the sensor dimension is intrinsic to various sensors. For instance, IMU sensor records three axises of accelerations. The WiFi sensor depends on the number of antennas and subcarriers. In this sense, we explore whether the aforementioned attacks to only a part of sensor dimensions can deteriorate the recognition models. For wireless sensing, we propose the Antenna-Level (AL) attack and Subcarrier-Level (SL) attack that perturbs the input sample in the antenna and subcarrier level, respectively. The attack perturbations are added to a single subcarrier or an antenna, and we will investigate whether such partial attack is harmful to the model. 
}

\subsection{Existing Solutions and Challenges}\label{sec:sub-challenges}
For device-free HAR systems such as wireless sensing as shown in Figure \ref{fig:intuition}, attackers can falsify the sensing data from two fragile points, the IoT system board or the communication link. Since the IoT security has always been a severe issue \cite{hassija2019survey}, attackers can hack into the IoT system and modify the data or the on-board model, which can incur white-box attacks, such as FGSM attack. Furthermore, when the data is transmitted to cloud server or other computational device, the communication link could be vulnerable due to some lightweight but insecure IoT communication protocol, thus vulnerable to black-box attacks, such as GN attack.

To defend FGSM or GN attacks, adversarial training is an effective approach that simulates the perturbations of attacks and augments the input data during model training. Though adversarial training has shown promising defense performance \cite{goodfellow2014explaining}, it still has the following limitations that make it hard to use for device-free HAR systems in the real world:
\begin{enumerate} 
	\item[1.] The hyper-parameters in GN and FGSM attacks are unknown for the defender. The attacker can build batches of different adversarial examples by simply changing the hyper-parameters, i.e. $\mu,\sigma$ in GN and $\epsilon$ in FGSM. However, the defender is not aware of these hyper-parameters for adversarial training, and maximizing the range of these hyper-parameters is quite time-consuming.
	\item[2.] Adversarial training brings difficulties to the convergence of model training. When minimizing the training loss of adversarial examples, it hinders the original task loss minimization, often leading to under fitting. This can be tackled by manual tuning of adversarial training, but this decreases the availability of such defense especially for unmanned systems.
	\item[3.] The defense capacity of adversarial training is quite limited, especially for the bimodal attack or AL attack. 
\end{enumerate}
To deal with these challenges summarized by observations of adversarial defense on wireless HAR, we develop a generic framework, SecureSense, which is able to defend GN, FGSM, bimodal and AL attacks, while converging stably without the need of manual hyper-parameter tuning.

\section{SecureSense: Improving Robustness for Wireless Sensing}\label{sec:defense}

As discussed in Section \ref{sec:attack}, device-free HAR systems are vulnerable to both white-box and black-box attacks. The objective of our SecureSense framework is to confer a significant reduction in a deep HAR model’s vulnerability to adversarial examples. Denote $h(x)$ as a deep HAR recognition model for wireless CSI data. In modern deep models, $h(x)$ consists of a feature extractor $f(x)$ and a classifier $g(x)$ parameterized by $\theta_f$ and $\theta_g$, respectively. 

\subsection{Supervised Representation Learning}
To enable the deep model $h(x)$ to recognize $K$ categories of human activities, it is indispensable to conduct supervised learning on normal sensing data. This is achieved by minimizing the cross-entropy loss
\begin{equation}
\mathcal{L}_{ce}(x,y)=-\mathbb{E}_{(x,y)}\sum_k \big[ \mathbb{I}[y=k] \log \big(\varphi(g(f(x))) \big) \big],
\end{equation}
where $\varphi(\cdot)$ is the softmax function, and $\mathbb{I}[y=k]$ means a 0-1 function that outputs 1 for the correct category $k$. By back-propagation, the feature extractor $f(x)$ can learn a discriminative latent space for $K$ categories, and the classifier $g(x)$ offers a good decision boundary in the feature space. From this perspective, adversarial examples are generated by perturbing the vulnerable examples near the decision boundary. In this sense, the perturbation makes a spurious sample by pushing it across the classification boundary, successfully deceiving the model $h(x)$.

\subsection{Adversarial Examples Generation}
Normal adversarial training for defense leverages adversarial examples as a way of data augmentation or feature augmentation, such as stability training \cite{zheng2016improving}. Deep neural network can thus increase its robustness by having seen these examples before and learned their adversarial distributions. According to the definition of GN and FGSM attacks, we construct the adversarial examples by
\begin{equation}
\tilde{x}_{GN} = x + \tau \Psi_{\mathcal{N}},
\end{equation}
\begin{equation}
	\tilde{x}_{FGSM} = x + \epsilon \text{sign} \big(\nabla_x L(x,y;\theta_h) \big),
\end{equation}
where $\tau,\epsilon$ denote the attack level, and $\tilde{x}_{GN}$ and $\tilde{x}_{FGSM}$ are the adversarial examples of GN and FGSM attack, respectively. Directly adding them into training dataset as augmentation can improve the model robustness, but it raises the challenges mentioned in Section~\ref{sec:sub-challenges}, rendering such augmentation hard to work well in practice. 

\subsection{Consistency Learning}
In our method, we enforce the model to generate consistent predictions for the original sample $x$ and its adversarial variations $\tilde{x}_{adv}$. Such objective helps model to adjust its decision boundary to defend adversarial attack. Denote $p(y|x;\theta_h)$ as the predicted probability of an input sample $x$ by the model $h(x)$, which is calculated by
\begin{equation}
	p(y|x;\theta_h) = \varphi(g(f(x))).
\end{equation}
Then for the original data sample $x_{ori}$ and its simulated adversarial examples $\tilde{x}_{adv}$, the model outputs their corresponding probabilities $p(y|x_{ori};\theta_h)$ and $p(y|\tilde{x}_{adv};\theta_h)$.

Considering $N_{adv}$ types of adversarial examples $\{\tilde{x}^j_{adv}\}_{j=1}^{N_{adv}}$, we denote their probabilities as $p_{adv}^j=p(y|\tilde{x}_{adv}^j;\theta_h)$, and the probability of the original sample as $p_{ori}=p(y|x_{ori};\theta_h)$. 
The SecureSense aims to generate consistent predictions. The simplest way is to minimize the cross-entropy loss between the ground truth label $y$ and all probabilities of adversarial examples, but this is hard to optimize due to the hard label. In empirical study, it is found that conflicts of training objectives may occur for the original and adversarial classification loss. A feasible method is to minimize the sum of the Kullback–Leibler (KL) divergences between each possible pairs of $p_{ori}$ and $p_{adv}^j$, which softly reduces the prediction disparity and is leveraged for model robustness enhancement such as adversarial logit pairing \cite{kannan2018adversarial}, 
The KL divergence between two probabilities $p(x)$ and $q(x)$ is calculated by
\begin{equation}
	D_{KL}(p||q)=\int_x p(x) \log\frac{p(x)}{q(x)}dx.
\end{equation}
Nevertheless, in this case, the number of KL loss terms will increase dramatically to $N_{adv}+{N_{adv} \choose 2}$ as the types of adversarial examples increase. It is difficult to optimize a number of KL losses, which also results in enormous computations and instable training procedure \cite{kannan2018adversarial}.  To deal with this problem, we employ the Jensen-Shannon (JS) Divergence \cite{hendrycks2019augmix} that aims to attain consistency between the mean probabilities and each probability.

Toward this end, we firstly compute the mean probabilities $\bar{p}$ across the original and adversarial examples by
\begin{equation}
	\bar{p}=\frac{1}{N_{adv}+1} \bigg(\sum_{j=1}^{N_{adv}}p_{adv}^j + p_{ori} \bigg).
\end{equation}
Then we calculate the JS divergence loss among the probabilities of all types of adversarial examples and the original example:
\begin{align}\nonumber
	&\mathcal{L}_{JS}(x_{ori},\tilde{x}_{adv}^1,\dots,\tilde{x}_{adv}^{N_{adv}})= \text{JS}(p_{ori},p_{adv}^1,\dots,p_{adv}^{N_{adv}})
	=\\
	&\frac{1}{N_{adv}+1}
	\bigg(D_{KL}(p_{ori}||\bar{p}) 
	+\sum_{j=1}^{N_{adv}}D_{KL}(p_{adv}^j||\bar{p})\bigg).
\end{align}
The number of loss terms is $N_{adv}+1$ and it can be trained effectively in a stable manner due to the usage of a smoother term $\bar{p}$.  By minimizing the JS divergence loss, our SecureSense is able to make consistent predictions for the original data sample and various adversarial examples. 

The overall learning objective of the SecureSense is formulated as
\begin{equation}\label{eq:total}
	\min_{\theta_f,\theta_g} \mathcal{L}_{ce}+\mathcal{L}_{JS}(x_{ori},\tilde{x}_{adv}^1,\dots,\tilde{x}_{adv}^{N_{adv}}).
\end{equation}
Two losses are jointly optimized by backpropagation as a multi-task learning. 

{\color{black}
\subsection{Algorithm Summary}
The training procedure of the SecureSense is summarized in Algorithm~\ref{algo:robustsense}. The SecureSense framework is a proactive defense algorithm that prevents potential attacks in the phase of model training. In SecureSense, we generate multiple kinds of adversarial examples by the vanilla model and leverages our consistency loss for augmentation. When our model achieves convergence, the model is robustness to adversarial perturbations. Empirically we find that the convergence is usually stable in the experiments. For a specific HAR application such as WiFi-based HAR, we consider FGSM and GN attacks in our algorithm. After initialization, we firstly calculate the gradients with respect to the original input by minimizing the cross-entropy loss. Then the gradients are used to generate FGSM adversarial examples, and GN adversarial examples are simulated by adding a Gaussian noise. Feeding these into the deep model, we can minimize the multi-task loss in Eq.(\ref{eq:total}) by backpropagation and eventually obtain the robust model parameters $\theta_h$. As the multi-task loss is an average of the symmetric KL-divergence, it has the same computational complexity as the cross-entropy for one sample. With FGSM and GN augmentation, the whole training procedure takes around three times as long as the vanilla training. After the model is trained, the SecureSense does not introduce any computations in the inference phase. It is noteworthy that the SecureSense can leverage existing adversarial attacks to defend unknown attack type (i.e. the proposed bimodal attack) and classic attacks with unknown parameters (i.e. $\epsilon,\mu,\sigma$) by means of consistency learning, which will be validated in the experiments.

}

\begin{algorithm}[t]
	\small
	\LinesNumbered
	\SetAlgoLined
	\SetAlgoLongEnd
	\DontPrintSemicolon
	\SetKwInput{KwModule}{Module}
	\SetKw{KwBegin}{BEGIN:}
	\SetKw{KwEnd}{END.}
	
	\small
	\caption{SecureSense Training\label{algo:robustsense}}
	
	\KwModule{
		a deep HAR model $h(x)$ composed of a feature extractor $f(x)$, a classifier $g(x)$, 
	}
	\KwIn{$x_{ori},\tilde{x}_{adv}^1,\dots,\tilde{x}_{adv}^{N_{adv}}$}
	
	\KwBegin{}\;
	Initialize $h(x)$ with $\theta_h$;\\: 
	\While(){epoch $<$ total epoch}{
		 Update $\theta_h$ by minimizing $\mathcal{L}_{ce}$\;
		
		$\tilde{x}_{FGSM} \gets x + \epsilon \text{sign} \big(\nabla_x L(x,y;\theta_h) \big)$\;
		
		$\tilde{x}_{GN} \gets x + \tau \Psi_{\mathcal{N}}$\;
		
		$p(y|x_{ori};\theta_h)\gets \sigma(g(f(x_{ori})))$\;
		$p(y|\tilde{x}_{FGSM};\theta_h)\gets \sigma(g(f(\tilde{x}_{FGSM})))$\;
		$p(y|\tilde{x}_{GN};\theta_h)\gets \sigma(g(f(\tilde{x}_{GN})))$\;
		$\bar{p}\gets (p(y|x_{ori};\theta_h)+p(y|\tilde{x}_{FGSM};\theta_h)+p(y|\tilde{x}_{GN};\theta_h))/3$\;
		
		Update $\theta_h$ by minimizing $\mathcal{L}_{ce}+\mathcal{L}_{JS}$
	}
	
	\KwOut{the model parameters $\theta_h$.}
	\KwEnd
\end{algorithm}

{\color{black}
\section{Experiment}\label{sec:experiment}
\subsection{Experiment Setup}
\textbf{System Design.} We build the wireless HAR system based on an IoT-enabled CSI sensing platform \cite{yang2018device} that consists of two TP-Link N750, one as a transmitter and the other as a receiver. Three pairs of antennas are leveraged and the system operates with 40MHz bandwidth on 5GHz. The Atheros CSI tool is able to report 114 subcarriers of CSI data for each timestamp \cite{xie2015precise}. We collect two datasets including a Human Activity Recognition (HAR) dataset and a gait dataset for Human Identification (HID) dataset. Each data sample is a matrix with a size of $3\times114\times500$. For the HAR dataset, six categories of human activities are collected including running, walking, falling down, boxing, circling arms, and cleaning floor. The sample number of each category is 400. The environment for data collection is shown in Figure \ref{fig:layout}. For the HID dataset, we collect the CSI gait samples of 12 volunteers. The volunteer is asked to walk through the Line-of-Sight (LoS) path of two WiFi devices from different angles and both directions, and we record 60 sample for each volunteer. Two datasets are split into the training and testing set by the ratio of 8:2.

\begin{table}[tp]
	\centering
	\begin{tabular}{c|c|c}
		\toprule \midrule
		Layer Index & Feature Extractor $f(x)$         & Label Predictor $g(x)$   \\ \midrule
		input           & \multicolumn{2}{c}{CSI data: 3 $\times$ 114 $\times$ 500}                         \\ \midrule
		1           & Conv 32$\times$(15,23), stride 9, ReLU            & 128 dense \\ \midrule
		2           & Conv 32$\times$(3,7), stride 1, ReLU     &   6 dense, softmax                \\ \midrule
		3           & Max-pool (1,2), stride (1,2)   &                   \\ \midrule
		4           & Conv 64$\times$(3,7), stride 1, ReLU    &                   \\ \midrule
		5           & Conv 96$\times$(3,7), stride 1, ReLU  &        \\ \midrule
		6           & Max-pool (1,2), stride (1,2)   & \\ \midrule \bottomrule
	\end{tabular}
	\caption{The network architecture used in the SecureSense experiments. For Conv A$\times$(H,W), A denotes the channel number, and (H,W) represents the height and width of the operation kernel. This applies to all Convolution (Conv) and Max-pooling (Max-pool) layers.}\label{table:network}
\end{table}

\begin{figure}[t]
	\centering
	\includegraphics[width=0.95\linewidth, angle=0]{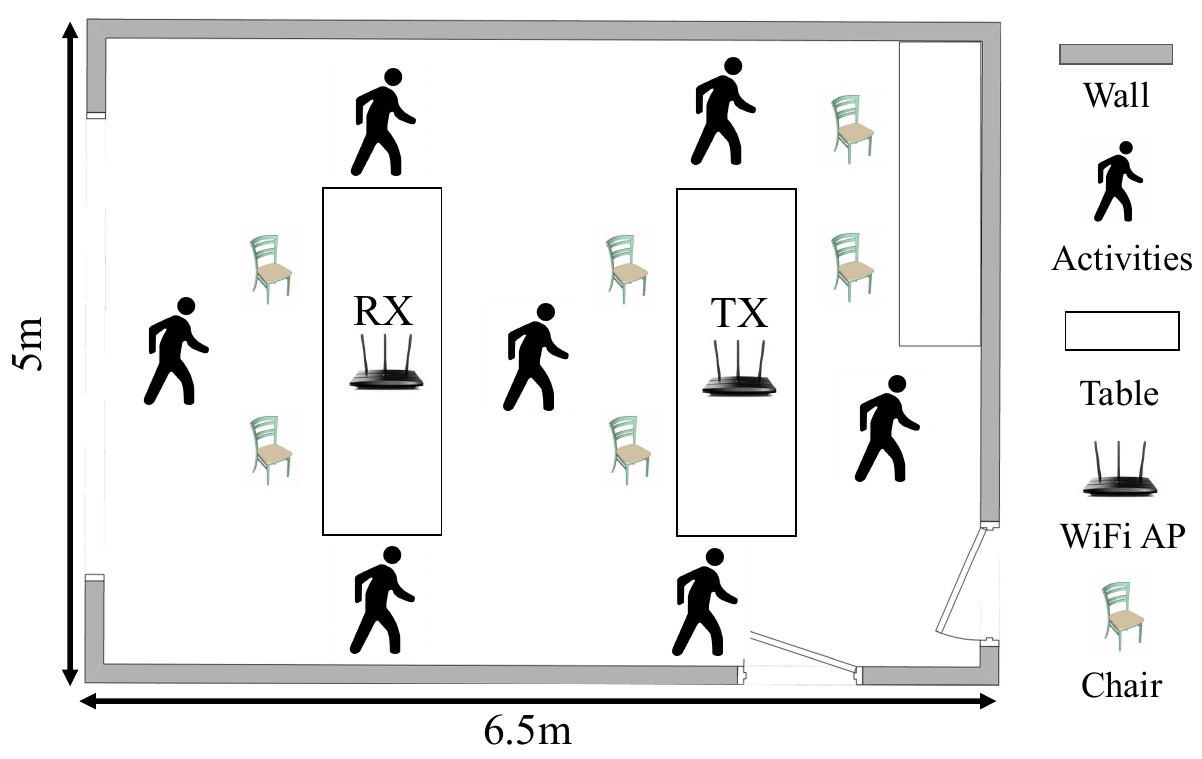}
	\caption{The layout of the experiment environment in the lab.}
	\label{fig:layout}
\end{figure}

\textbf{Implementation Details.} The network design is shown in Table~\ref{table:network} where the feature extractor $f(x)$ is a convolutional neural network and the classifier $g(x)$ only consists of fully-connected layers. Such a typical network can achieve high accuracy on the proposed dataset, which is sufficient for our evaluation under adversarial attack. We prototype the SecureSense on a server with two NVIDIA RTX 2080Ti GPUs for evaluation. The PyTorch framework is utilized for implementation. The details of various attacks have been introduced in Section~\ref{subsec:attack}. For the hyper-parameters $\epsilon,\mu,\sigma$, they are specified for different settings of evaluation in the rest of the paper. All the networks are optimized by a mini-batch Adam with a learning rate of 0.01 and a momentum of 0.9. We apply a total epoch of 100 and a batch size of 128 to guarantee sufficient training iterations for the HAR and HID tasks.

\textbf{Evaluation Metrics and Baselines.} To compare the model performance under different levels of adversarial attacks, we report the classification accuracy of the model under the $m$-th setting of the $n$-th type of attack as $E^n_m$, and also report the mean accuracy cross all settings as $\bar{E}^n=\frac{1}{m}\sum_{i=1}^m E^n_i$ that evaluates the model robustness to a specific attack. We compare our method with the original model without any augmentation, regarded as the baseline model, and the adversarial training methods based on different augmentation losses \cite{goodfellow2014explaining,kannan2018adversarial}, Defense-GAN \cite{samangouei2018defense}, MagNet \cite{meng2017magnet}, and Collaborative Multi-Task Training (CMT) \cite{wang2020CMT}. In the experiment tables, we denote AdvT as the adversarial training, $\mathcal{L}_y,\mathcal{L}_{G},\mathcal{L}_{F}$ as the cross-entropy loss in terms of the original sample, GN examples and FGSM examples. The adversarial training works with either or both of $\mathcal{L}_{G}$ and $\mathcal{L}_{F}$. The Defense-GAN leverages the powerful generative adversarial network to model the unperturbed distributions and eliminates adversarial changes from adversarial examples \cite{samangouei2018defense}. The MagNet captures the manifold of normal examples and move adversarial samples to the normal manifolds \cite{meng2017magnet}. The CMT incorporates label pairs into feature learning and deals with various attacks successfully \cite{wang2020CMT}. Since our method is also based on adversarial training, we compare SecureSense with all settings of AdvT on all tasks, and other baselines are evaluated on HAR tasks.

\begin{table}[t]
	\centering
	\caption{Accuracy (\%) on HAR dataset under GN attack}\label{tb:gn-har}
	\scalebox{0.89}{
		\begin{tabular}{l|ccccccc|c}
			\toprule
			\multirow{2}{*}{Method} & \multicolumn{7}{c|}{$\tau$} & \multirow{2}{*}{Mean} \\ 
			& 0 & 0.1 & 0.2 & 0.3 & 0.5 & 1.0 & 2.0 & \\ \midrule
			Baseline & 96.6 & 95.1 & 94.7 & 93.6 & 91.3 & 76.9 & 51.9 & 85.7 \\
			AdvT ($\mathcal{L}_y,\mathcal{L}_{G}$) & 93.9 & 95.8 & 94.7 & 94.7 & 94.7 & 93.9 & 92.4 & 94.3 \\
			AdvT (All) & 95.8 & 95.8 & 94.7 & 95.8 & 93.9 & 86.0 & 79.5 & 91.6 \\
			AdvT ($\mathcal{L}_{G},\mathcal{L}_{F}$) & 91.7 & 92.8 & 93.2 & 93.6 & 93.9 & 94.3 & 93.9 & 93.3 \\
			AdvT ($\mathcal{L}_y,\mathcal{L}_{F}$) & \textbf{98.1} & 98.1 & 92.8 & 87.1 & 68.9 & 51.9 & 44.3 & 77.3 \\
			MagNet \cite{meng2017magnet} & 96.7 & 95.3 & 95.8 & 94.7 & 93.9 & 93.9 & 92.4 & 94.7 \\
			CMT \cite{wang2020CMT} & 96.3 & 95.8 & 94.7 & 96.8 & 94.7 & 89.1 & 81.7 & 92.7 \\
			Defense-GAN \cite{samangouei2018defense} & 96.6 & 97.1 & 98.1 & 94.7 & 98.1 & 98.1 & 94.7 & 96.8 \\
			SecureSense & \textbf{98.1} & \textbf{98.9} & \textbf{99.2} & \textbf{99.6} & \textbf{99.6} & \textbf{99.6} & \textbf{95.8} & \textbf{98.7} \\ \bottomrule
		\end{tabular}
	}
\end{table}

\begin{table}[t]
	\centering
	\caption{Accuracy (\%) on HAR dataset under FGSM attack}\label{tb:fgsm-har}
	\scalebox{0.89}{
		\begin{tabular}{l|ccccccc|c}
			\toprule
			\multirow{2}{*}{Method} & \multicolumn{7}{c|}{$\epsilon$} & \multirow{2}{*}{Mean} \\
			& 0 & 0.05 & 0.1 & 0.15 & 0.2 & 0.25 & 0.3 &  \\ \midrule
			Baseline & 96.6 & 71.2 & 35.6 & 14.8 & 9.8 & 5.3 & 4.5 & 34 \\
			AdvT ($\mathcal{L}_y,\mathcal{L}_{G}$) & 93.9 & 76.5 & 28 & 10.2 & 7.6 & 7.6 & 7.6 & 33.1 \\
			AdvT (All) & 95.8 & 80.3 & 55.7 & 46.2 & 37.1 & 32.2 & 29.5 & 53.8 \\
			AdvT ($\mathcal{L}_{G},\mathcal{L}_{F}$) & 91.7 & 89.8 & 86.7 & 81.1 & 72.3 & 65.5 & 61.7 & 78.4 \\
			AdvT ($\mathcal{L}_y,\mathcal{L}_{F}$) & \textbf{98.1} & 98.5 & 93.9 & 87.9 & 78.8 & 68.2 & 58.3 & 83.4 \\
			MagNet \cite{meng2017magnet} & 96.7 & 83.1 & 60.1 & 57.1 & 46.2 & 65.5 & 61.7 & 67.2 \\
			CMT \cite{wang2020CMT} & 96.3 & 92.8 & 93.9 & 83.1 & 75.7 & 68.2 & 67.7 & 82.5 \\
			Defense-GAN \cite{samangouei2018defense} & 96.6 & 93.6 & 93.2 & 89.8 & 86.7 & 86.7 & 81.1 & 89.7 \\
			SecureSense & \textbf{98.1} & \textbf{98.9} & \textbf{98.5} & \textbf{95.1} & \textbf{92.4} & \textbf{85.2} & \textbf{80.3} & \textbf{92.6} \\ \bottomrule
		\end{tabular}
	}
\end{table}

\begin{table}[t]
	\centering
	\caption{Accuracy (\%) on HAR dataset under bimodal attack}\label{tb:bimodal-har}
	\scalebox{0.89}{
		\begin{tabular}{l|ccccccc|c}
			\toprule
			\multirow{2}{*}{Method} & \multicolumn{7}{c|}{Attack Level} & \multirow{2}{*}{Mean} \\ 
			& 0 & 1 & 2 & 3 & 4 & 5 & 6 &  \\ \midrule
			Baseline & 96.6 & 79.9 & 38.3 & 15.2 & 6.1 & 7.2 & 15.9 & 37 \\
			AdvT ($\mathcal{L}_y,\mathcal{L}_{G}$) & 93.9 & 75.8 & 31.8 & 13.3 & 9.5 & 8.0 & 5.3 & 33.9 \\
			AdvT (All) & 95.8 & 81.8 & 66.3 & 54.9 & 50.0 & 60.6 & 57.2 & 66.6 \\
			AdvT ($\mathcal{L}_{G},\mathcal{L}_{F}$) & 91.7 & 89.8 & 87.9 & 81.4 & 76.5 & 68.9 & 66.3 & 80.4 \\
			AdvT ($\mathcal{L}_y,\mathcal{L}_{F}$) & 98.1 & 93.6 & 82.2 & 66.7 & 53 & 39.8 & 36.7 & 67.2 \\
			MagNet \cite{meng2017magnet} & 96.7 & 83.8 & 82.2 & 66.7 & 65.5 & 63.7 & 57.2 & 73.7 \\
			CMT \cite{wang2020CMT} & 96.3 & 93.6 & 93.2 & 75.7 & 67.7 & 68.9 & 66.3 & 80.2 \\
			Defense-GAN \cite{samangouei2018defense} & 96.6 & 93.6 & 89.8 & 86.7 & 75.7 & 76.5 & 50.0 & 81.3 \\
			SecureSense & \textbf{98.1} & \textbf{99.2} & \textbf{98.5} & \textbf{96.2} & \textbf{88.3} & \textbf{81.4} & \textbf{84.5} & \textbf{92.3} \\ \bottomrule
		\end{tabular}
	}
\end{table}

\subsection{Overall Evaluation}
\subsubsection{Evaluation on HAR dataset}
We firstly evaluate the proposed method and compare it with the adversarial training methods on HAR dataset. The evaluation here is to fix the noise weights $\tau,\epsilon$ for training and testing. For the GN attack, the noise weight $\tau$ is set to $[0,0.1,0.2,0.3,0.5,1.0,2.0]$, while for the FGSM attack, the weight $\epsilon$ is set to $[0,0.05,0.1,0.15,0.2,0.25,0.3]$. For the bimodal attack, we set 7 attack levels based on the ordered combination of the range list of $\tau$ and $\epsilon$. As shown in Table~\ref{tb:gn-har}, the SecureSense outperforms all other methods, achieving an amazing average accuracy of 98.7\%. As the noise level $\tau$ increases, the accuracies of all methods decrease, but our method still shows strong performance with nearly no negative effect. Such decreasing degree is even worse for FGSM attack, as shown in Table~\ref{tb:fgsm-har}. With a small $\epsilon=0.1$, the accuracy of the baseline HAR model drops from 96.6\% to 35.6\%. The FGSM attack can even deteriorate the model performance to an extremely low value that is worse than the random guess, i.e. 4.3\% for the baseline model at $\epsilon=0.3$. To make matters worse, adversarial training seems not to work well. It is observed that AdvT with $\mathcal{L}_y,\mathcal{L}_{G}$ can help improve the model performance to 94.3\% for evaluation under GN attack, but this does not apply to the FGSM situation. For AdvT with $\mathcal{L}_y,\mathcal{L}_{F}$, it only achieves $58.3\%$ for FGSM attack with $\epsilon=0.3$. In comparison, the proposed SecureSense significantly surpasses all other methods, achieving a mean accuracy of 92.6\%. Though the state-of-the-art defense methods show improvements, our method still outperforms MagNet, CMT and Defense-GAN by 25.4\%, 10.1\% and 3.1\%, respectively. For the bimodal attack as shown in Table~\ref{tb:bimodal-har}, most of the AdvT methods fail to defend the high-level attacks. It is obvious that the mixture of both attacks brings more challenges to the deep HAR models. Our method still gains a superior accuracy of 92.3\% in average, surpassing the second place model by 11.0\%. Furthermore, for the most difficult case at ``attack level 6'', our model can achieve an accuracy of 84.5\%, outperforming the best comparative model (CMT) by around 18\%. It is observed that the performance of Defense-GAN decreases by 8.4\% under bimodal attack, which indicates that GAN-based defense approach may not overcome mixtures of attack categories. Whereas, our method and CMT still achieve consistent results in this challenging scenario.
}

\begin{table}[t]
	\centering
	\caption{Accuracy (\%) on HID dataset under GN attack}\label{tb:gn-human-id}
	\scalebox{0.89}{
		\begin{tabular}{l|ccccccc|c}
			\toprule
			\multirow{2}{*}{Method} & \multicolumn{7}{c|}{$\tau$} & \multirow{2}{*}{Mean} \\ 
			& 0 & 0.1 & 0.2 & 0.3 & 0.5 & 1.0 & 2.0 & \\ \midrule
			Baseline & 91.2 & 90.8 & 90.3 & 89.2 & 87.7 & 79.5 & 66.3 & 85 \\
			AdvT ($\mathcal{L}_y,\mathcal{L}_{G}$) & 78.4 & 76.7 & 72.2 & 68.9 & 54.9 & 25.6 & 11.4 & 55.4 \\
			AdvT (All) & 90.1 & 87.4 & 86.3 & 85.5 & 81.3 & 68.1 & 61.7 & 80.1 \\
			AdvT ($\mathcal{L}_{G},\mathcal{L}_{F}$) & 82.1 & 77.7 & 72.3 & 64.8 & 51.1 & 28.6 & 10.4 & 55.3 \\
			AdvT ($\mathcal{L}_y,\mathcal{L}_{F}$) & 91 & 91.9 & 92.5 & 92.5 & 91.0 & 84.8 & 68.5 & 87.5 \\
			SecureSense & \textbf{95.8} & \textbf{95.4} & \textbf{95.1} & \textbf{94.5} & \textbf{93.4} & \textbf{93.4} & \textbf{92.1} & \textbf{94.2} \\ \bottomrule
		\end{tabular}
	}
\end{table}

\begin{table}[t]
	\centering
	\caption{Accuracy (\%) on HID dataset under FGSM attack}\label{tb:fgsm-human-id}
	\scalebox{0.89}{
		\begin{tabular}{l|ccccccc|c}
			\toprule
			\multirow{2}{*}{Method} & \multicolumn{7}{c|}{$\epsilon$} & \multirow{2}{*}{Mean} \\
			& 0 & 0.05 & 0.1 & 0.15 & 0.2 & 0.25 & 0.3 &  \\ \midrule
			Baseline & 91.2 & 65 & 43.6 & 23.8 & 14.3 & 7.7 & 4.6 & 35.7 \\
			AdvT ($\mathcal{L}_y,\mathcal{L}_{G}$) & 78.4 & 72.3 & 63.7 & 53.3 & 41.4 & 32.1 & 23.3 & 52.1 \\
			AdvT (All) & 90.1 & 86.8 & 75.1 & 60.1 & 43.2 & 32.8 & 24 & 58.9 \\
			AdvT ($\mathcal{L}_{G},\mathcal{L}_{F}$) & 82.1 & 79.5 & 70.5 & 56.4 & 45.2 & 33.7 & 23.3 & 55.8 \\
			AdvT ($\mathcal{L}_y,\mathcal{L}_{F}$) & 91.0 & 61.7 & 39.2 & 26 & 16.7 & 11.4 & 8.2 & 36.3 \\
			SecureSense & \textbf{95.8} & \textbf{92.5} & \textbf{89.0} & \textbf{82.6} & \textbf{77.1} & \textbf{70.1} & \textbf{62.6} & \textbf{81.4} \\ \bottomrule
		\end{tabular}		
	}
\end{table}

\begin{table}[t]
	\centering
	\caption{Accuracy (\%) on HID dataset under bimodal attack}\label{tb:bimodal-human-id}
	\scalebox{0.89}{
		\begin{tabular}{l|ccccccc|c}
			\toprule
			\multirow{2}{*}{Method} & \multicolumn{7}{c|}{Attack Level} & \multirow{2}{*}{Mean} \\ 
			& 0 & 1 & 2 & 3 & 4 & 5 & 6 &  \\ \midrule
			Baseline & 91.2 & 65.2 & 44.1 & 25.1 & 15.8 & 10.1 & 7.7 & 37 \\
			AdvT ($\mathcal{L}_y,\mathcal{L}_{G}$) & 78.4 & 72.7 & 62.8 & 49.8 & 33.5 & 17.8 & 10.8 & 46.5 \\
			AdvT (All) & 90.1 & 86.6 & 80 & 65.9 & 48.9 & 35.7 & 20.9 & 61.2 \\
			AdvT ($\mathcal{L}_{G},\mathcal{L}_{F}$) & 82.1 & 74.4 & 61.9 & 50.7 & 34.8 & 19.2 & 9 & 47.4 \\
			AdvT ($\mathcal{L}_y,\mathcal{L}_{F}$) & 91 & 60.1 & 39 & 24.7 & 16.7 & 11.5 & 9.5 & 36.1 \\
			SecureSense & \textbf{95.8} & \textbf{92.3} & \textbf{88.8} & \textbf{81.9} & \textbf{76.9} & \textbf{70.5} & \textbf{60.6} & \textbf{81.0} \\ \bottomrule
		\end{tabular}	
	}
\end{table}

\begin{table}[thp]
	\centering
	\caption{Accuracy (\%) on HAR dataset under three types of attacks by random training and testing.}\label{tb:random-har}
	\scalebox{0.92}{
		\begin{tabular}{l|l|ccccc|c}
			\toprule
			\multirow{2}{*}{Attack} & \multirow{2}{*}{Method} & \multicolumn{5}{c|}{Runs} & \multirow{2}{*}{Mean} \\ 
			&  & 1 & 2 & 3 & 4 & 5 &  \\ \midrule
			\multirow{6}{*}{\begin{tabular}[c]{@{}l@{}}GN\\ Attack\end{tabular}} & Baseline & 76.9 & 75.8 & 77.7 & 78 & 76.9 & 77.1 \\
			& AdvT ($\mathcal{L}_y,\mathcal{L}_{G}$) & 94.7 & 93.6 & 93.6 & 89.8 & 92.8 & 92.9 \\
			& AdvT (All) & 87.1 & 87.1 & 88.6 & 89.4 & 88.6 & 88.2 \\
			& AdvT ($\mathcal{L}_{G},\mathcal{L}_{F}$) & 39.4 & 38.3 & 34.8 & 37.9 & 33 & 36.7 \\
			& AdvT ($\mathcal{L}_y,\mathcal{L}_{F}$) & 93.2 & 96.6 & 93.2 & 96.2 & 92.8 & 94.4 \\
			& SecureSense & \textbf{98.5} & \textbf{97.7} & \textbf{98.5} & \textbf{98.9} & \textbf{98.5} & \textbf{98.4} \\ \midrule
			\multirow{6}{*}{\begin{tabular}[c]{@{}l@{}}FGSM\\ Attack\end{tabular}} & Baseline & 33.7 & 30.7 & 28.4 & 26.1 & 31.4 & 30.1 \\
			& AdvT ($\mathcal{L}_y,\mathcal{L}_{G}$) & 82.6 & 84.1 & 83.3 & 84.8 & 85.6 & 84.1 \\
			& AdvT (All) & 81.8 & 83.7 & 83 & 81.1 & 81.8 & 82.3 \\
			& AdvT ($\mathcal{L}_{G},\mathcal{L}_{F}$) & 79.9 & 80.7 & 79.2 & 79.2 & 76.9 & 79.2 \\
			& AdvT ($\mathcal{L}_y,\mathcal{L}_{F}$) & 41.3 & 46.6 & 42.4 & 42.8 & 43.2 & 43.3 \\
			& SecureSense & \textbf{94.7} & \textbf{93.6} & \textbf{93.9} & \textbf{93.2} & \textbf{94.7} & \textbf{94} \\ \midrule
			\multirow{6}{*}{\begin{tabular}[c]{@{}l@{}}Bimodal\\ Attack\end{tabular}} & Baseline & 37.5 & 34.5 & 29.2 & 31.1 & 34.1 & 33.3 \\
			& AdvT ($\mathcal{L}_y,\mathcal{L}_{G}$) & 79.2 & 77.7 & 75.4 & 77.7 & 77.3 & 77.5 \\
			& AdvT (All) & 78.4 & 76.1 & 76.9 & 78.4 & 75.8 & 77.1 \\
			& AdvT ($\mathcal{L}_{G},\mathcal{L}_{F}$) & 34.1 & 38.3 & 33 & 37.1 & 38.3 & 36.2 \\
			& AdvT ($\mathcal{L}_y,\mathcal{L}_{F}$) & 48.5 & 45.8 & 50.4 & 43.2 & 46.2 & 46.8 \\
			& SecureSense & \textbf{93.6} & \textbf{93.2} & \textbf{93.6} & \textbf{92} & \textbf{91.7} & \textbf{92.8} \\
			\bottomrule
		\end{tabular}
	}
\end{table}

\subsubsection{Evaluation on HID dataset}\label{sec:hid-ev}
Based on the same setting, we further evaluate the model on HID dataset with more classes but less training samples. The gait activities across people are similar, which leads to more difficulties. The results are alike to those of HAR dataset for FGSM and bimodal attack. The SecureSense attains the state-of-the-art performances as shown in \Cref{tb:fgsm-human-id,tb:bimodal-human-id,tb:gn-human-id}. Furthermore, it is demonstrated that the SecureSense achieves more stable and consistent results for all noise level $\tau,\sigma$, while the AdvT methods perform poorly for large noise levels. Compared to the results on HAR dataset, the improvement margin increases, which indicates that our SecureSense has better robustness for tough datasets with more activity classes.

\subsubsection{Findings and Discussions}
We also have some interesting findings based on the evaluations on two datasets and multiple experimental settings:
\begin{itemize}
	\item[1.] The SecureSense not only enhances the capacity of model defense, but also improves the vanilla model performance without attacks. It is seen that the SecureSense shows better results than the baseline models when $\tau=0$ and $\epsilon=0$. Such incremental effect also applies to adversarial training. Learning robustness to perturbations helps model to seek for a better classifier boundary.
	
	\item[2.] The normal adversarial training cannot stably converge, especially for the case based on GN examples. In \Cref{tb:fgsm-human-id,tb:bimodal-human-id,tb:gn-human-id}, the AdvT ($\mathcal{L}_y,\mathcal{L}_G$) achieves 78.4\% for HID without attack, less than the baseline of 91.2\% by a large margin. In comparison, the SecureSense can always converge well.
	
	\item[3.] In adversarial training, the best performance does not stem from the case of employing all adversarial examples, i.e. AdvT (All). This might be caused by the fact that the optimization of three cross-entropy losses is quite difficult, which is detrimental to the final performance. Overfitting to adverarial examples is also harmful, and the conclusion we draw is quite similar to the research on Projected Gradient Descent (PGD) \cite{madry2017towards}. In contrast, our method leverages all adversarial examples and yields the best results with a stable training procedure.
	
	\item[4.] In adversarial training, the usage of FGSM samples can effectively help defend GN attack, but the contrary situation does not work. 
\end{itemize}

\begin{figure*}[htp]
	\centering
	\subfigure[AdvT with $\mathcal{L}_y,\mathcal{L}_G,\mathcal{L}_F$]{
		\includegraphics[width=0.48\textwidth]{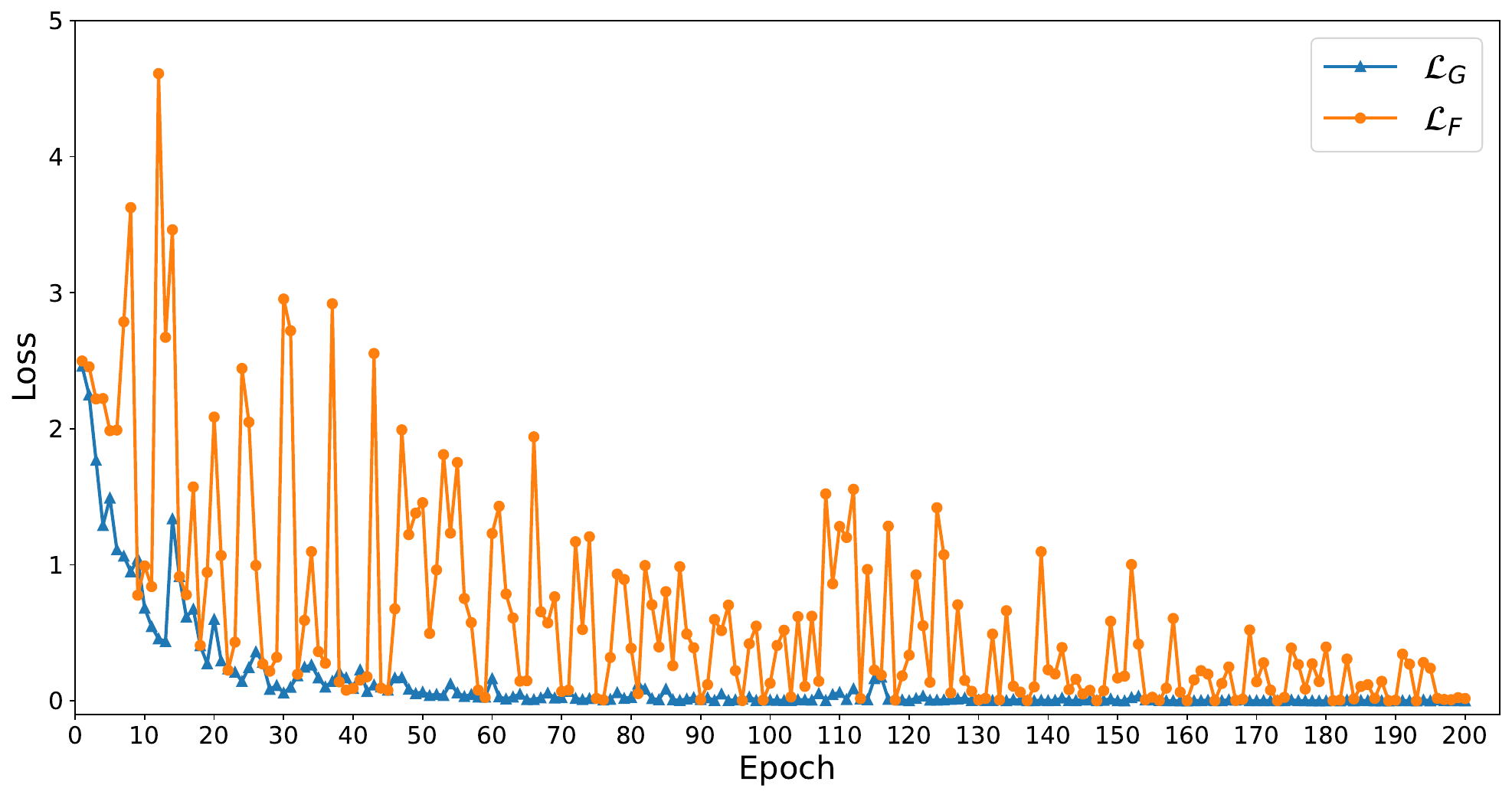}
	}
	\subfigure[SecureSense]{
		\includegraphics[width=0.48\textwidth]{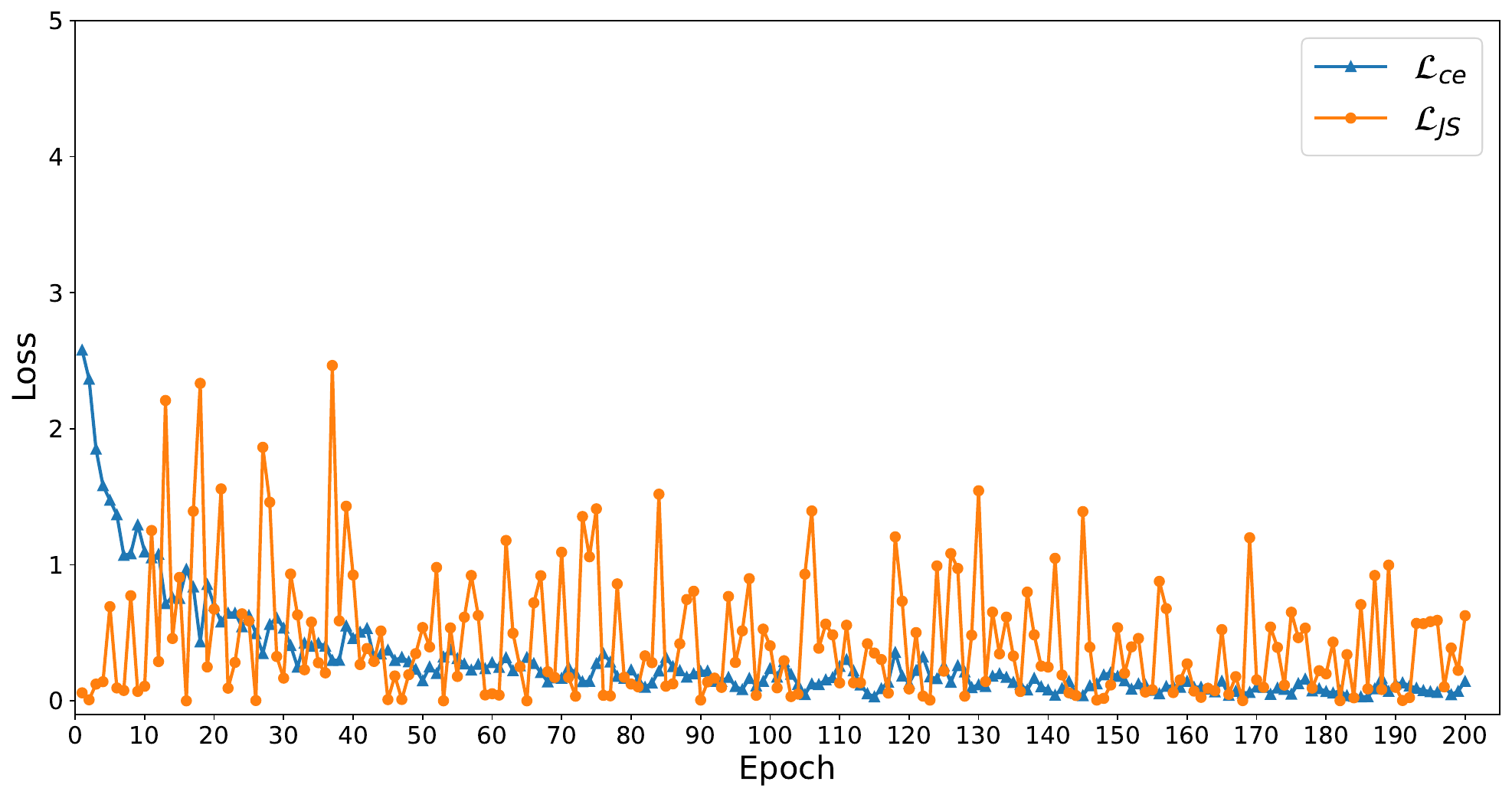}}\\
	\caption{The training losses of the AdvT and SecureSense.}
	\label{fig:training-loss}
\end{figure*}

\begin{table}[tp]
	\centering
	\caption{Accuracy (\%) on HID dataset under FGSM attack at the antenna and subcarrier level.}\label{tb:local-attack-hid}
	\scalebox{0.78}{
		\begin{tabular}{l|l|cccccll|c}
			\toprule
			\multirow{2}{*}{Attack} & \multirow{2}{*}{Method} & \multicolumn{7}{c|}{$\epsilon$} & \multirow{2}{*}{Mean} \\ 
			&  & 0 & 0.05 & 0.1 & 0.15 & 0.2 & 0.25 & 0.3 &  \\ \midrule
			\multirow{3}{*}{\begin{tabular}[c]{@{}l@{}}Subcarrier\\ Attack\end{tabular}} & Baseline & 90.8 & 90.8 & 90.7 & 90.7 & 90.7 & 90.5 & 90.3 & 90.6 \\
			& AdvT ($\mathcal{L}_{G},\mathcal{L}_{F}$) & 89.2 & 89.2 & 89.2 & 89.2 & 89.2 & 89.2 & 88.8 & 89.1 \\
			& SecureSense & \textbf{95.8} & \textbf{95.8} & \textbf{95.8} & \textbf{95.8} & \textbf{95.8} & \textbf{95.8} & \textbf{95.8} & \textbf{95.8} \\ \midrule
			\multirow{3}{*}{\begin{tabular}[c]{@{}l@{}}Antenna\\ Attack\end{tabular}} & Baseline & 90.8 & 85.9 & 80 & 74 & 64.3 & 57.5 & 49.5 & 71.7 \\
			& AdvT ($\mathcal{L}_{G},\mathcal{L}_{F}$) & 89.2 & 83.9 & 79.5 & 75.3 & 71.6 & 69.6 & 68.5 & 76.8 \\
			& SecureSense & \textbf{95.8} & \textbf{94.0} & \textbf{92.9} & \textbf{91.0} & \textbf{88.5} & \textbf{85.7} & \textbf{82.4} & \textbf{90.0} \\ 
			\bottomrule
		\end{tabular}
	}
\end{table}

\subsection{Evaluation on Random Attack Cases}\label{sec:random-attack}
The fixed number and values of hyper-parameters $\tau,\epsilon$ in defense and attack might not conform with the real world cases, since defender cannot know these settings defined by attacker, and attacker can specify multiple distinct settings for adversarial attack. In this sense, we simulate this situation by a more challenging evaluation. In this experiment, the $\tau$ and $\epsilon$ for defender and attacker are randomly generated for each training epoch and each testing attack run. This is to say -- the hyper-parameters $\tau,\epsilon$ for training the model is absolutely different from those for the testing phase.

We employ a continuous uniform distribution $U(0,2)$ and $U(0,0.3)$ to generate random $\tau$ and $\epsilon$, respectively. Extensive experiments are conducted on the HAR dataset, and the results across five runs with random seeds are shown in Table~\ref{tb:random-har}. It is seen that the realistic attacks are harder to tackle, leading to more degrading performance to baselines and adversarial training approaches. In Table~\ref{tb:gn-har}, most models achieves over 80\% accuracy for the situation under GN attack, but the performance gets worse for the random $\tau$. The similar phenomena are observed for FGSM attack. For bimodal attack, we can see that the SecureSense attains a mean accuracy of 92.8\%, outperforming the second place AdvT by over 15\%. These results demonstrate that our approach is robust to complicated real-world attacks.

\subsection{Evaluation on Local Attack}
We also investigate the proposed local attack at the antenna or subcarrier level. It is found that the local GN attack is too weak, so we only conduct experiments on AL and SL attack via FGSM adversarial examples. All the models are trained as the same setting as Section~\ref{sec:hid-ev}. As shown in Table~\ref{tb:local-attack-hid}, the SL attack does not obviously decrease the model performance. The reason is that the subcarrier-level perturbation is too small to threat the model. If we enlarge this perturbation to the antenna level, it is shown that the baseline has a decreasing performance of 71.7\%. The best AdvT method improves it by 5.1\%, while our method achieves a better accuracy of 84.2\%, which demonstrates the superiority of the SecureSense. Apart from the performance comparison, we also come to a conclusion from the attacker perspective: multi-view HAR system is more robust to local attacks. Even though sensors are partially contaminated, deep models can still make correct predictions based on clean ones. For example, there exist massive subcarrier sensors in wireless sensing, and each subcarrier gives a spatial description of human activities. In this fashion, multiple subcarriers offer multi-view data, which is robust to local attack.

\begin{table}
	\centering
	\begin{tabular}{|m{1.6cm}<{\centering}|m{6cm}<{\centering}|}
		\toprule
		Method & Samples \\ \midrule
		Raw examples & \includegraphics[width=0.28\textwidth]{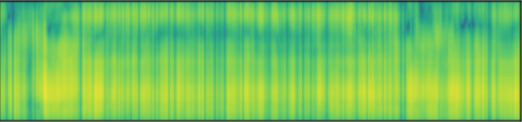} \\ \midrule
		GN attack & \includegraphics[width=0.28\textwidth]{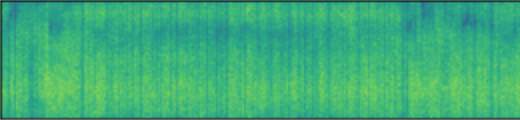} \\ \midrule
		FGSM attack & \includegraphics[width=0.28\textwidth]{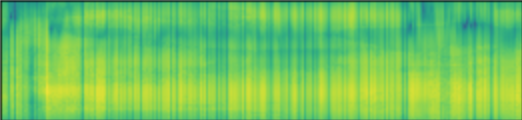} \\ \midrule
		Bimodal attack & \includegraphics[width=0.28\textwidth]{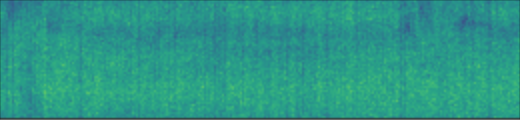}
		\\ \bottomrule
	\end{tabular}
	\caption{Raw and adversarial WiFi CSI examples.}\label{tb:data-example}
\end{table}

\subsection{Analytics}
\subsubsection{Training Stability}\label{sec:stability}
To evaluate the training stability, we use SGD and Adam optimizer with an initial learning rate ranging from 0.1 to 0.001. We discover that the SecureSense can stably converge across multiple runs, but the AdvT cannot. Actually, to obtain the results for the previous comparison, we make efforts to tune the hyper-parameter, i.e. the trade-off between $\mathcal{L}_G,\mathcal{L}_F$. In Figure~\ref{fig:training-loss}, we visualize the training losses of our method and a stable case of the AdvT. It is discovered that the cross-entropy loss of FGSM examples are hard to optimize due to the fact that FGSM examples are generated for each epoch dynamically. Such variances of loss may lay a negative impact on the optimization of other losses, hindering the convergence. In contrast, the JS loss in our method has lower values than $\mathcal{L}_F$, and it does not affect the optimization procedure empirically.

\subsubsection{Attack Sample Visualization}\label{sec:exp-attack-vis}
We visualize the raw data and the perturbed data of the same example in HID dataset as shown in Table~\ref{tb:data-example}. In the human view, the example under GN attack is obviously different from the raw example, and the situation is worse for the one under bimodal attack. The FGSM example cannot be distinguished from the raw example due to the slight changes induced by FGSM and a super small $\epsilon$. However, the FGSM example is quite harmful to model performance. The visualization reminds us that simply comparing two examples cannot be a good defense strategy for adversarial attack.

\begin{figure*}[tp]
	\centering
	\includegraphics[width=0.24\textwidth]{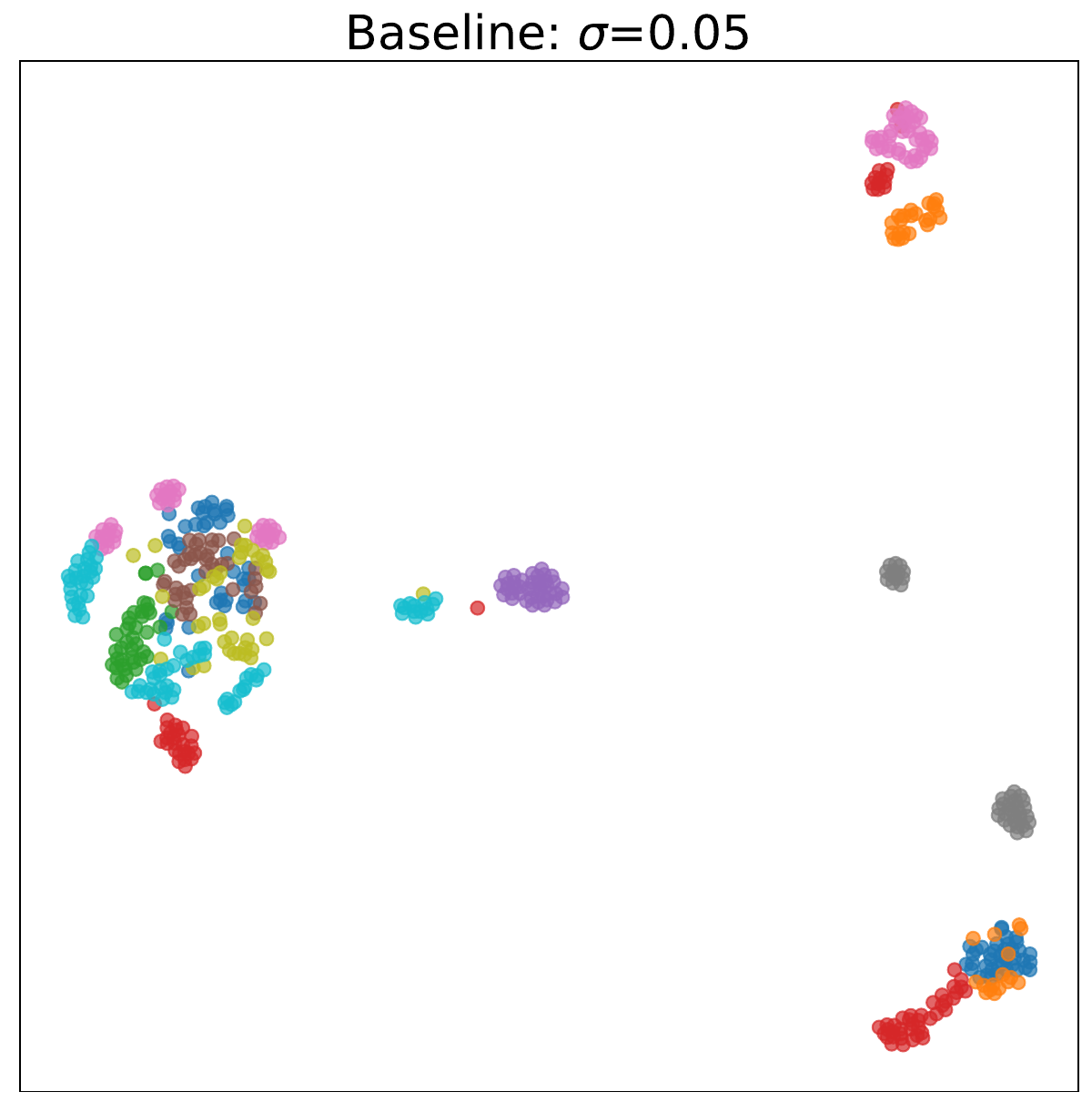}
	\includegraphics[width=0.24\textwidth]{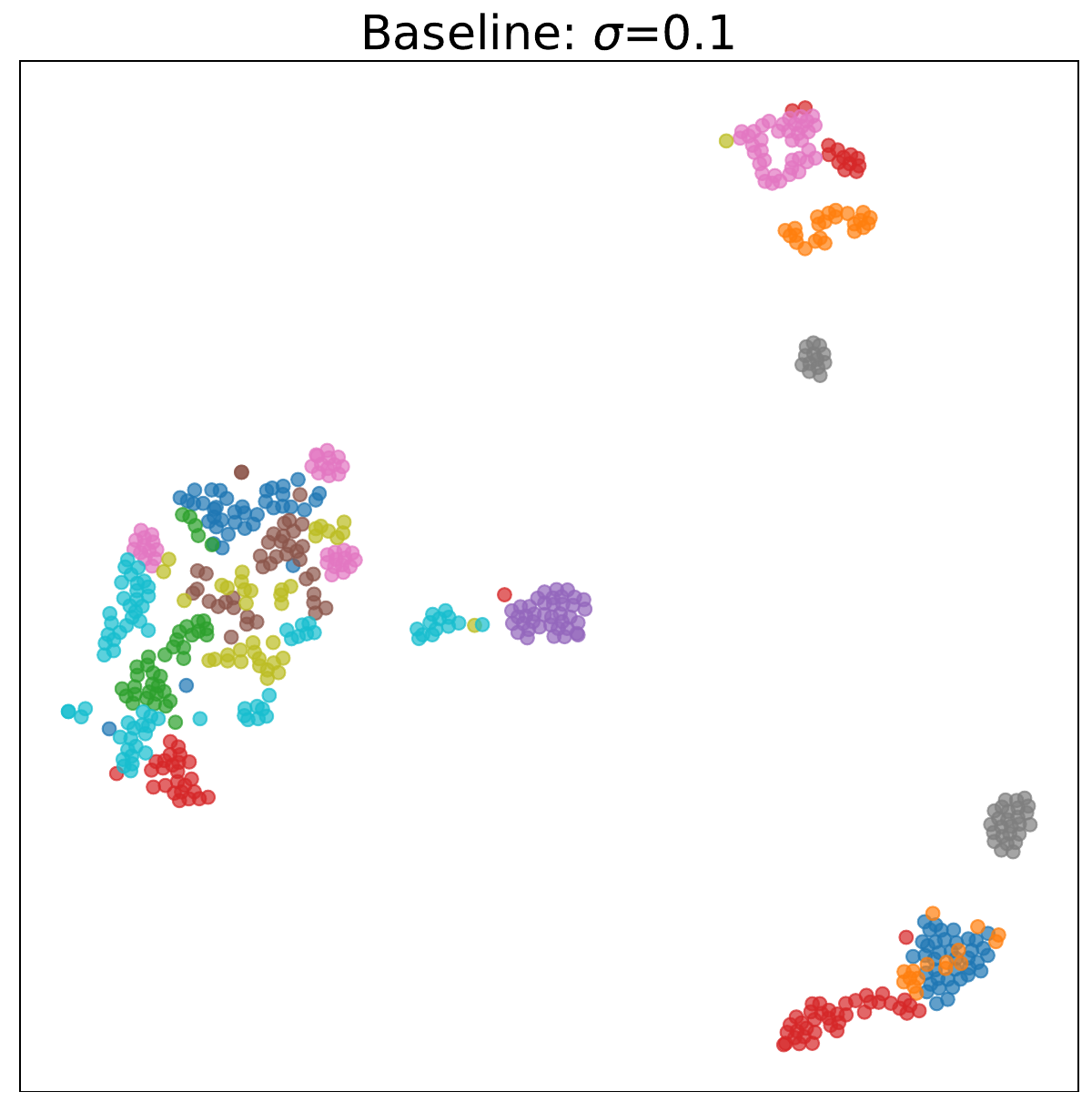}
	\includegraphics[width=0.24\textwidth]{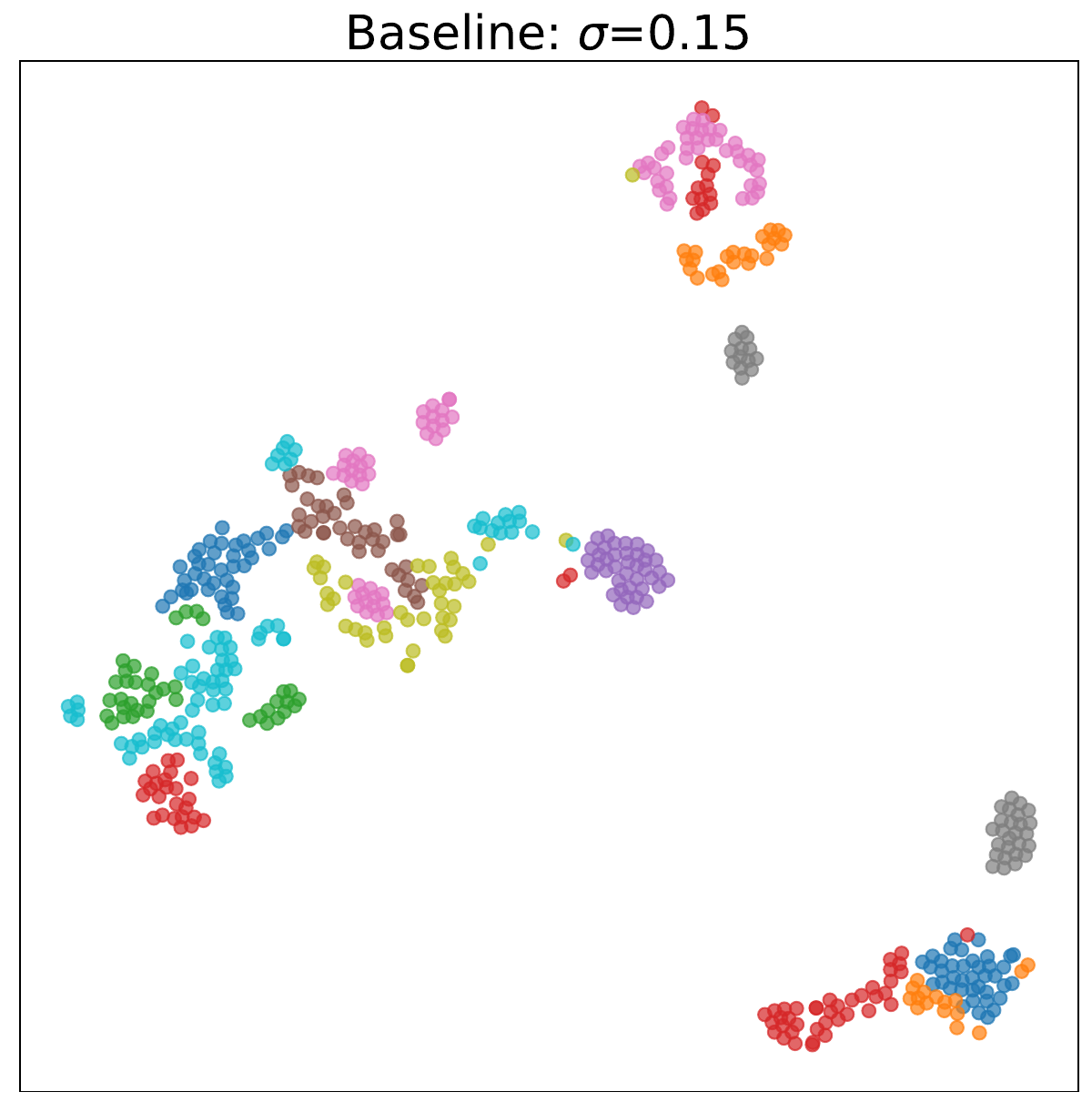}
	\includegraphics[width=0.24\textwidth]{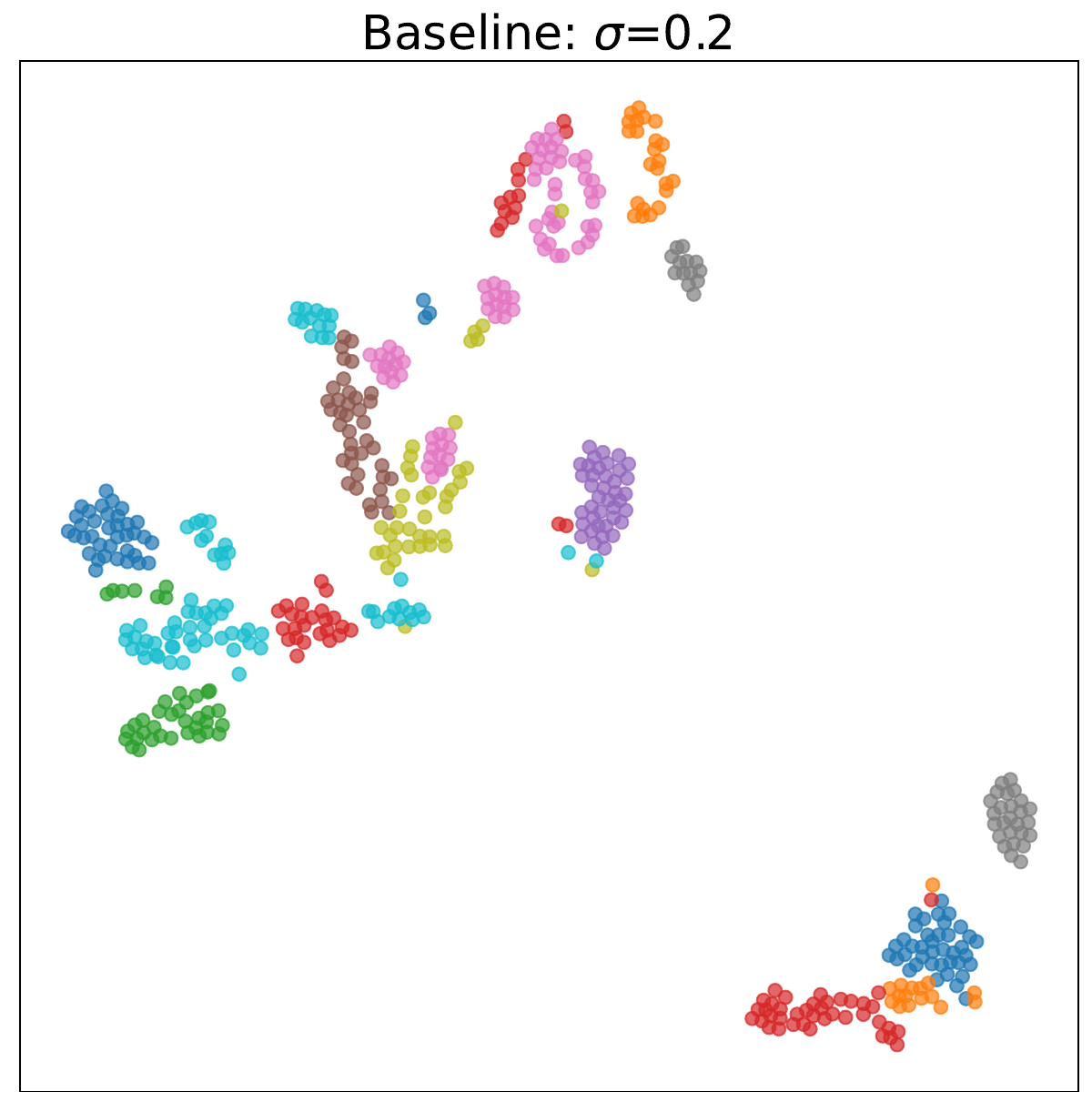}\\
	\includegraphics[width=0.24\textwidth]{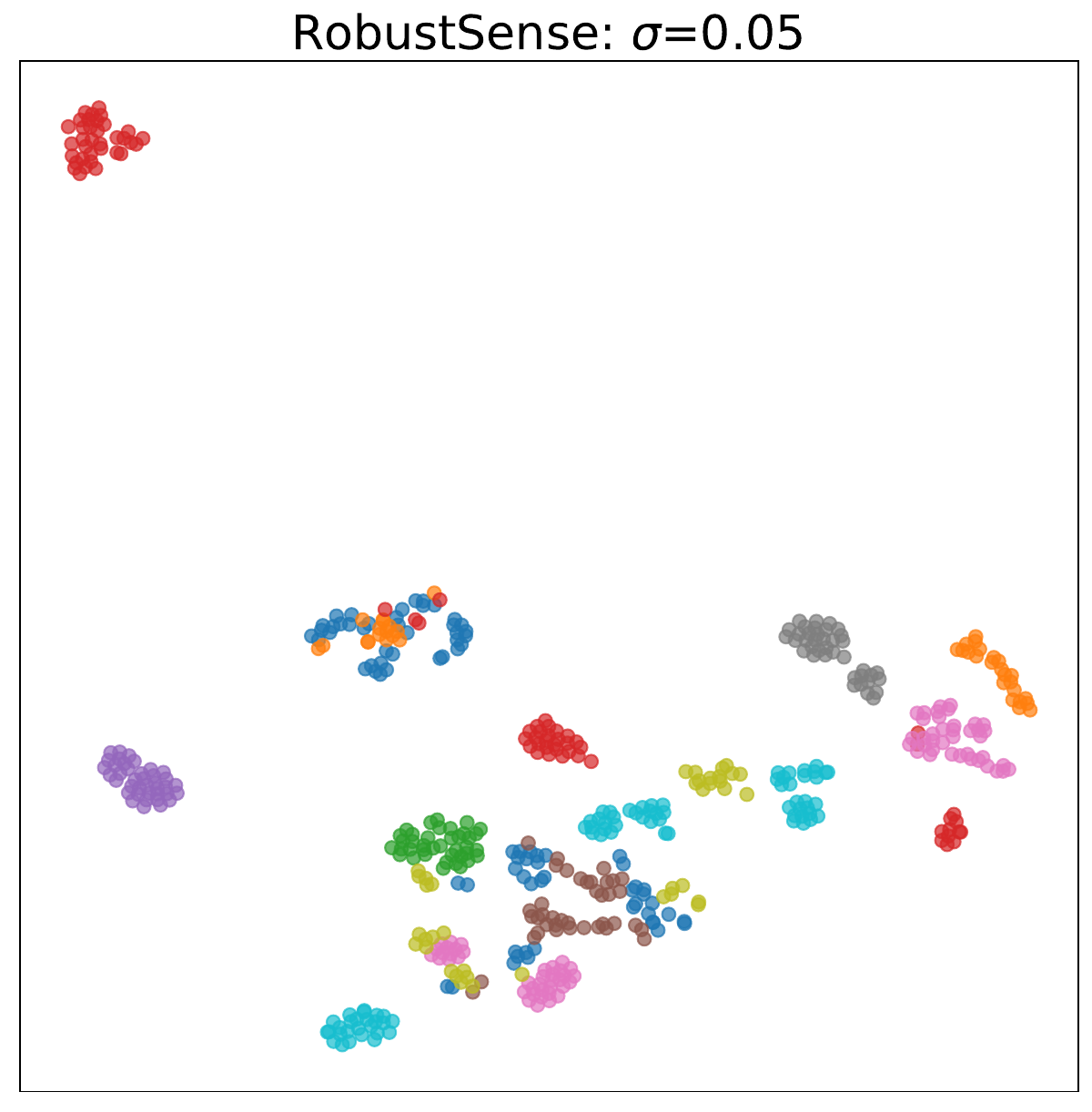}
	\includegraphics[width=0.24\textwidth]{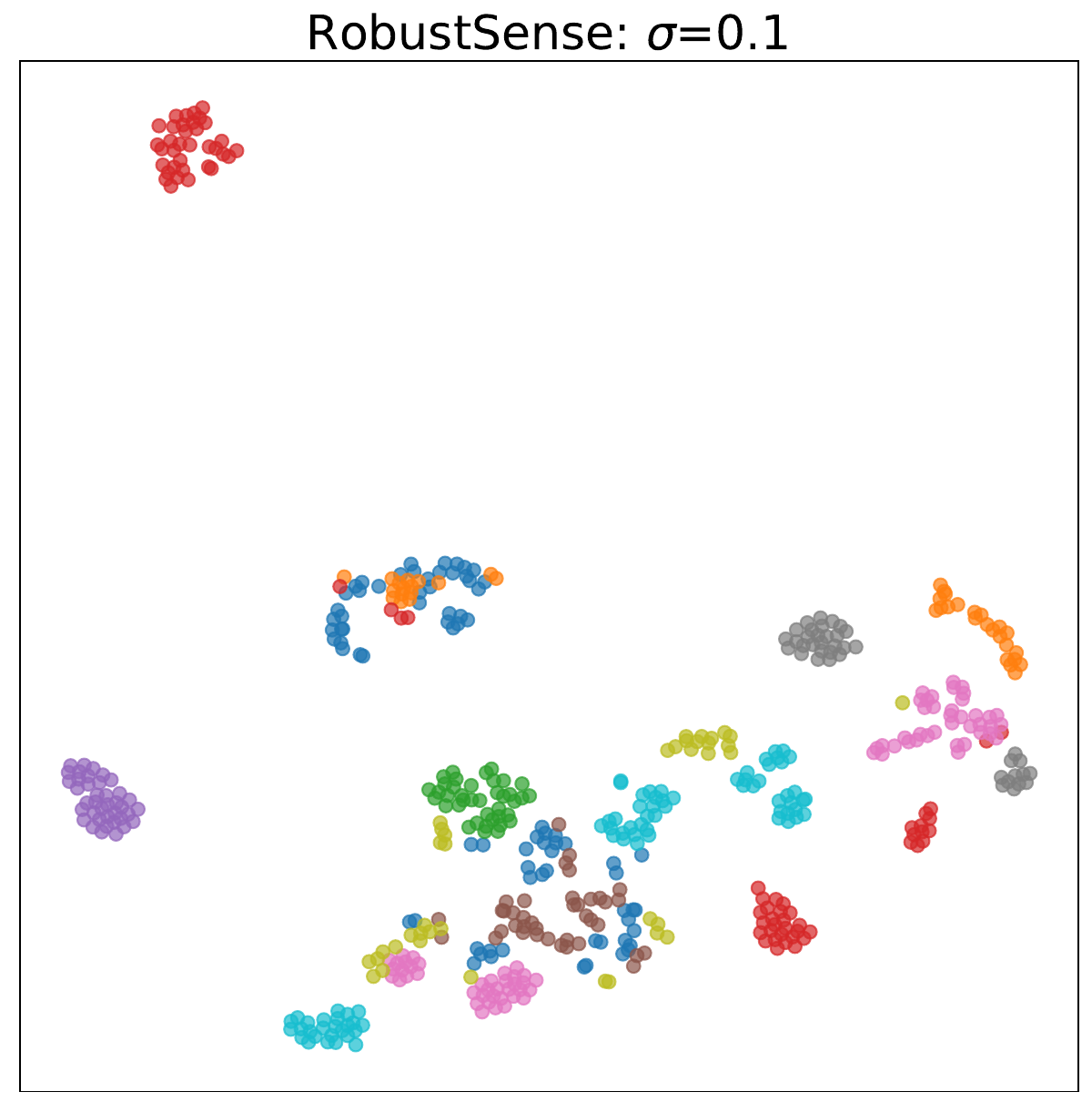}
	\includegraphics[width=0.24\textwidth]{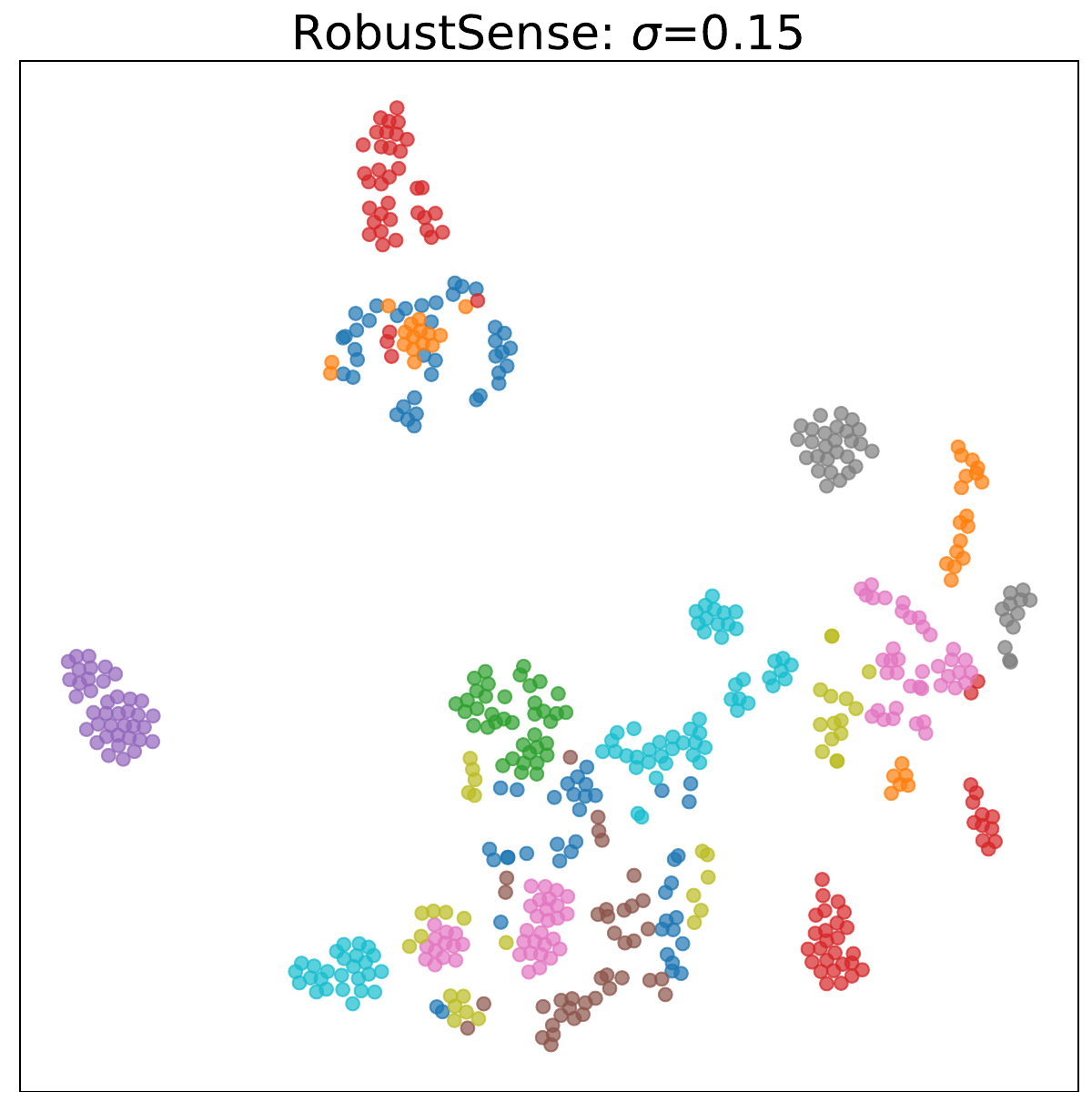}
	\includegraphics[width=0.24\textwidth]{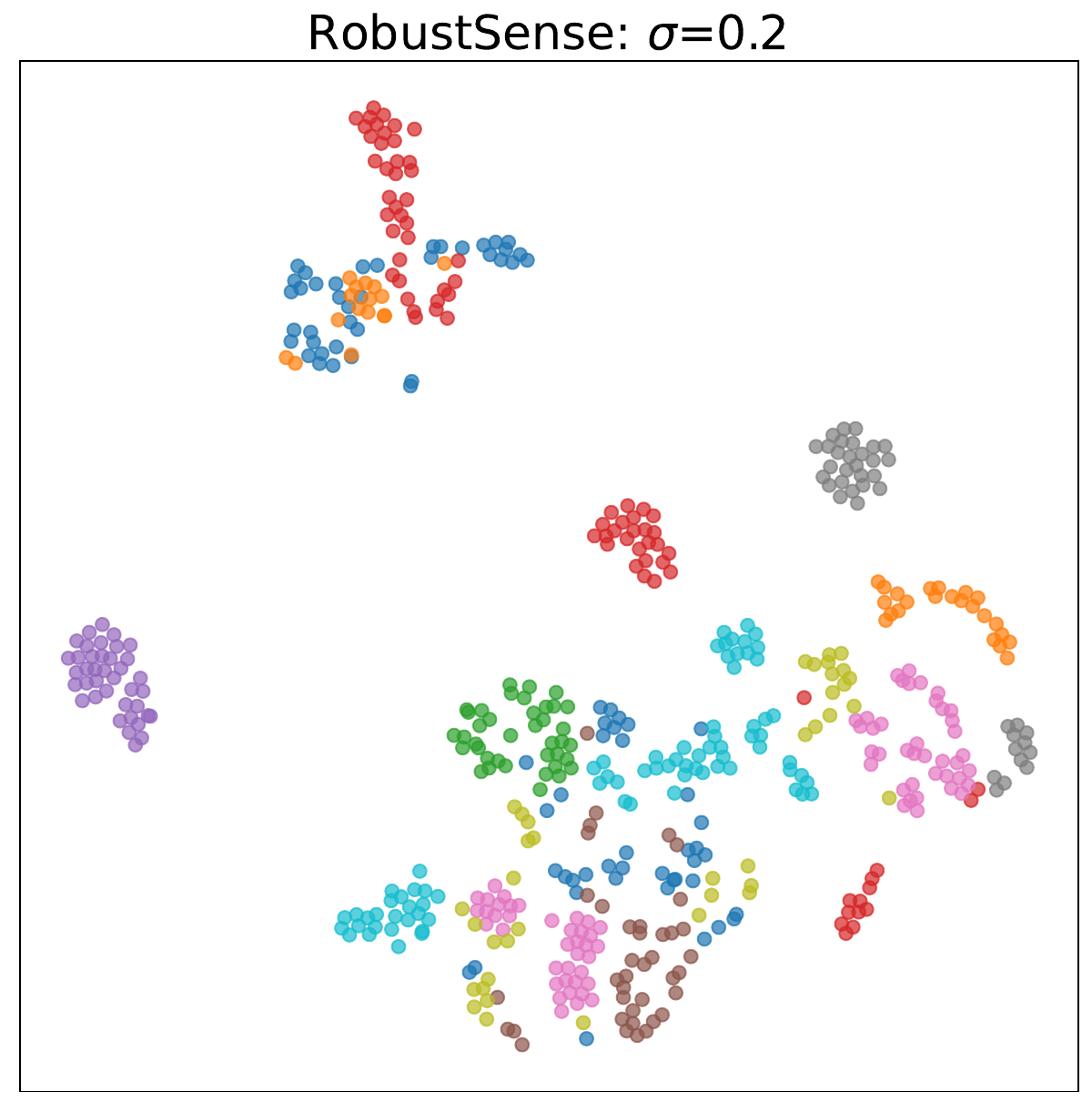}
	\caption{The t-SNE visualization of the latent space under FGSM attacks (1st row: baseline model; 2nd row: SecureSense). Each color indicates a category of human gaits. It is seen that the latent space of our method has a smaller degree of distortion. }
	\label{fig:tsne}
\end{figure*}

\subsubsection{Manifold Visualization}\label{sec:exp-tsne}
To study how adversarial attacks affect deep models, we apply t-SNE \cite{van2008visualizing} to the feature space of the baseline model and the SecureSense. As shown in Figure~\ref{fig:tsne}, the top and bottom figures show the feature space of the baseline and our method in terms of different attack levels. The FGSM attacks are conducted, and the four columns represent $\epsilon=0.05,0.1,0.15,0.2$, respectively. It is seen that the FGSM attacks can confuse the feature learning. As $\sigma$ increases, the samples of the same class become separate, and their clusters get loose. For the baseline method without any defense, there does not exist sufficient margin for the boundary between two classes. Compared to the baseline, the SecureSense can still preserve important structures of samples in the manifold, which allows the classifier to learn a robust decision boundary.

{\color{black}
\subsection{Discussions}
Based on the findings, we can discover how the proposed approach addresses the three challenges in Section~\ref{sec:sub-challenges}:
\begin{itemize}
	\item \textbf{Our method generalize well under attacks with unknown settings, addressing the hyper-parameter issue.} Based on the results in Section~\ref{sec:random-attack}, the proposed method achieves superior results for 5 independent runs with unknown hyper-parameters of three types of attacks. This validates that the SecureSense is robust to the hyper-parameters in the three attack types. In comparison, the adversarial training methods fail in most random attacks.
	
	\item \textbf{The SecureSense can easily achieve convergence with the novel consistency loss.} As shown in Figure~\ref{fig:training-loss} and demonstrated in all tables, the training of our method is smooth with fast convergence, and the improvements are consistent for various settings and datasets.
	
	\item \textbf{The SecureSense enhances the defense capacity even in hard situations such as mixtures of attacks.} The evaluation results show that the comparative methods show decreasing improvement in hard situations, i.e. under bimodal attack. Whereas, our approach still brings robust improvement, showing the better defense capacity to adversarial examples.
\end{itemize}
}

\section{Related Work}\label{sec:related-work}
\subsection{Device-Free HAR Systems}
Device-free HAR systems leverage various environment sensors to passively collect human motion patterns, including \cite{li2019survey}, WiFi \cite{wang2017device,yang2018device,zou2018deepsense,zou2018identification,zou2018joint,wang2022caution,yang2022metafi}, LED light \cite{nguyen2018eyelight} and ultrasound \cite{jiang2018towards}. They overcome the shortcomings of device-based HAR systems, e.g. IMU, since device-free HAR systems do not require users to carry a specific sensors. Despite convenience, HAR via device-free sensors is more difficult, as the surrounding environment may change continuously and bring about noises that hinder the HAR task \cite{zou2018robust}. Thanks to the progress of deep learning, subtle patterns can be extracted and recognized even under difficult circumstances \cite{hussain2020review}. Chen et al. propose an attentive recurrent network to recognize human activities based on CSI. Shi et al. propose a matching network for robust HAR, trained with only one sample of each activity \cite{shi2020environment}. Yang et al. reduce the communication cost by CSI compression~\cite{yang2022efficientfi} and enable data-efficient training by self-supervised learning for WiFi sensing~\cite{yang2022autofi}. Environment dependencies are removed by domain-invariant training \cite{zou2018robust,jiang2018towards}. The environment-independent HAR model can be compressed to work at the edge device via MobileDA \cite{yang2020mobileda}. Apart from deep models, signal processing approaches are good at detecting periodic vital sign or gait activities, e.g. heartbeat and respiration \cite{wang2016human,zhang2021wi,hu2022resfi}. 

{\color{black}
\subsection{Security of HAR Systems}
Regardless of HAR models, device-free HAR systems are vulnerable to adversarial attacks . The sensing data can be maliciously tampered in the IoT embedded system or the communication link to the cloud computation center \cite{luo2020attackers}. For example, Sagduyu et al. present over-the-air spectrum poisoning attacks for the data transmission \cite{sagduyu2019adversarial}. On the other hand, attackers can induce a small perturbation to sensing data that fails existing deep recognition models, and even misleads the model to a specific class \cite{goodfellow2014explaining}. The adversarial attacks have been widely explored in computer vision and natural language processing, e.g. face recognition \cite{dong2019efficient} and text classification \cite{li2020bert}. However, seldom research can be found on adversarial attacks for device-free HAR data. Existing approaches either add them to normal training (e.g. cross-entropy loss) \cite{madry2017towards} for data augmentation or design extra networks to eliminate adversarial parts from adverarial examples \cite{samangouei2018defense}. Whereas, our approach aims to learn consistency between adversarial and normal examples, which is validated to show superior performance.  Recently, Yang et al. demonstrate that Doppler-based HAR systems \cite{yang2020adversarial} are vulnerable to basic adversarial attack, which also obstructs the WiFi-based HAR systems \cite{huang2021wars}. Since the data properties of device-free HAR sensors (e.g. WiFi CSI data) are quite different from images, how adversarial attacks make an impact on HAR model and how to defend the negative effect is a non-trivial task. Compared to these works, we provide a comprehensive study on white-box and black-box data attacks for device-free wireless HAR systems. Besides, we firstly propose a defense framework that empowers HAR models to make robust predictions under Gaussian and FGSM attacks.

\section{Conclusion}\label{sec:conclusion}
In this paper, we study adversarial attack and defense for IoT-based HAR systems. Regarding the WiFi device-free HAR system as a case study, we propose several types of attacks including both white-box and black-box ones. The empirical results show that adversarial attacks can result in a degrading performance of deep models working in HAR systems. To enhance the capacity of HAR models confronting attacks, we propose a novel method, SecureSense, which aims to learn consistent predictions for normal cases and the hazardous cases under various attacks. Extensive experiments have been conducted, and the results validate that our approach achieves robust performances on two real-world human activity and gait recognition datasets. Visualization further reveals the underlying intuition behind adversarial attacks. 

The current work aims to enhance the model robustness to prevent potential attacks, which belongs to proactive defense. Meanwhile, when attacks happen, the model is reckoned to recognize the attacks for record, termed as reactive defense. It is still challenging to distinguish the adversarial samples from normal samples in HAR. In the future, we will study reactive defense methods for robust HAR systems that can reject adversarial samples from attackers, and more kinds of adversarial attacks will be considered. Meanwhile, apart from attacks in data communication, the attacks during data collection can also happen. For example, when we collect a dataset for human activity recognition, some volunteers may conduct wrong actions or irregular actions. In this case, the training data is noisy or consists of diverse distributions. Such problems could be solved by noise-robust learning \cite{peng2020suppressing} or domain adaptation \cite{yang2020mobileda}, which will also denote our future works.
}



\bibliographystyle{IEEEtran}
\bibliography{egbib}

\begin{thebibliography}{10}
\providecommand{\url}[1]{#1}
\csname url@samestyle\endcsname
\providecommand{\newblock}{\relax}
\providecommand{\bibinfo}[2]{#2}
\providecommand{\BIBentrySTDinterwordspacing}{\spaceskip=0pt\relax}
\providecommand{\BIBentryALTinterwordstretchfactor}{4}
\providecommand{\BIBentryALTinterwordspacing}{\spaceskip=\fontdimen2\font plus
\BIBentryALTinterwordstretchfactor\fontdimen3\font minus
  \fontdimen4\font\relax}
\providecommand{\BIBforeignlanguage}[2]{{%
\expandafter\ifx\csname l@#1\endcsname\relax
\typeout{** WARNING: IEEEtran.bst: No hyphenation pattern has been}%
\typeout{** loaded for the language `#1'. Using the pattern for}%
\typeout{** the default language instead.}%
\else
\language=\csname l@#1\endcsname
\fi
#2}}
\providecommand{\BIBdecl}{\relax}
\BIBdecl

\bibitem{chen2017robust}
Z.~Chen, Q.~Zhu, Y.~C. Soh, and L.~Zhang, ``Robust human activity recognition
  using smartphone sensors via ct-pca and online svm,'' \emph{IEEE Transactions
  on Industrial Informatics}, vol.~13, no.~6, pp. 3070--3080, 2017.

\bibitem{li2019survey}
X.~Li, Y.~He, and X.~Jing, ``A survey of deep learning-based human activity
  recognition in radar,'' \emph{Remote Sensing}, vol.~11, no.~9, p. 1068, 2019.

\bibitem{yang2018device}
J.~Yang, H.~Zou, H.~Jiang, and L.~Xie, ``Device-free occupant activity sensing
  using wifi-enabled iot devices for smart homes,'' \emph{IEEE Internet of
  Things Journal}, vol.~5, no.~5, pp. 3991--4002, 2018.

\bibitem{jiang2018towards}
W.~Jiang, C.~Miao, F.~Ma, S.~Yao, Y.~Wang, Y.~Yuan, H.~Xue, C.~Song, X.~Ma,
  D.~Koutsonikolas \emph{et~al.}, ``Towards environment independent device free
  human activity recognition,'' in \emph{Proceedings of the 24th Annual
  International Conference on Mobile Computing and Networking}, 2018, pp.
  289--304.

\bibitem{zhang2021gaitsense}
Y.~Zhang, Y.~Zheng, G.~Zhang, K.~Qian, C.~Qian, and Z.~Yang, ``Gaitsense:
  towards ubiquitous gait-based human identification with wi-fi,'' \emph{ACM
  Transactions on Sensor Networks (TOSN)}, vol.~18, no.~1, pp. 1--24, 2021.

\bibitem{yang2019learning}
J.~Yang, H.~Zou, Y.~Zhou, and L.~Xie, ``Learning gestures from wifi: A siamese
  recurrent convolutional architecture,'' \emph{IEEE Internet of Things
  Journal}, vol.~6, no.~6, pp. 10\,763--10\,772, 2019.

\bibitem{yang2018carefi}
J.~Yang, H.~Zou, H.~Jiang, and L.~Xie, ``Carefi: Sedentary behavior monitoring
  system via commodity wifi infrastructures,'' \emph{IEEE Transactions on
  Vehicular Technology}, vol.~67, no.~8, pp. 7620--7629, 2018.

\bibitem{yang2018fine}
------, ``Fine-grained adaptive location-independent activity recognition using
  commodity wifi,'' in \emph{2018 IEEE Wireless Communications and Networking
  Conference (WCNC)}.\hskip 1em plus 0.5em minus 0.4em\relax IEEE, 2018, pp.
  1--6.

\bibitem{deng2022gaitfi}
L.~Deng, J.~Yang, S.~Yuan, H.~Zou, C.~X. Lu, and L.~Xie, ``Gaitfi: Robust
  device-free human identification via wifi and vision multimodal learning,''
  \emph{arXiv preprint arXiv:2208.14326}, 2022.

\bibitem{szegedy2013intriguing}
C.~Szegedy, W.~Zaremba, I.~Sutskever, J.~Bruna, D.~Erhan, I.~Goodfellow, and
  R.~Fergus, ``Intriguing properties of neural networks,'' \emph{arXiv preprint
  arXiv:1312.6199}, 2013.

\bibitem{madry2017towards}
A.~Madry, A.~Makelov, L.~Schmidt, D.~Tsipras, and A.~Vladu, ``Towards deep
  learning models resistant to adversarial attacks,'' \emph{arXiv preprint
  arXiv:1706.06083}, 2017.

\bibitem{akhtar2018threat}
N.~Akhtar and A.~Mian, ``Threat of adversarial attacks on deep learning in
  computer vision: A survey,'' \emph{Ieee Access}, vol.~6, pp.
  14\,410--14\,430, 2018.

\bibitem{zhao2021expressive}
X.~Zhao, Z.~Zhang, Z.~Zhang, L.~Wu, J.~Jin, Y.~Zhou, R.~Jin, D.~Dou, and
  D.~Yan, ``Expressive 1-lipschitz neural networks for robust multiple graph
  learning against adversarial attacks,'' in \emph{International Conference on
  Machine Learning}.\hskip 1em plus 0.5em minus 0.4em\relax PMLR, 2021, pp.
  12\,719--12\,735.

\bibitem{goodfellow2014explaining}
I.~J. Goodfellow, J.~Shlens, and C.~Szegedy, ``Explaining and harnessing
  adversarial examples,'' \emph{arXiv preprint arXiv:1412.6572}, 2014.

\bibitem{yuan2019adversarial}
X.~Yuan, P.~He, Q.~Zhu, and X.~Li, ``Adversarial examples: Attacks and defenses
  for deep learning,'' \emph{IEEE transactions on neural networks and learning
  systems}, vol.~30, no.~9, pp. 2805--2824, 2019.

\bibitem{zhang2020adversarial}
W.~E. Zhang, Q.~Z. Sheng, A.~Alhazmi, and C.~Li, ``Adversarial attacks on
  deep-learning models in natural language processing: A survey,'' \emph{ACM
  Transactions on Intelligent Systems and Technology (TIST)}, vol.~11, no.~3,
  pp. 1--41, 2020.

\bibitem{xie2015precise}
Y.~Xie, Z.~Li, and M.~Li, ``Precise power delay profiling with commodity
  wifi,'' in \emph{Proceedings of the 21st Annual International Conference on
  Mobile Computing and Networking}.\hskip 1em plus 0.5em minus 0.4em\relax ACM,
  2015, pp. 53--64.

\bibitem{yang2022deep}
J.~Yang, X.~Chen, D.~Wang, H.~Zou, C.~X. Lu, S.~Sun, and L.~Xie, ``Deep
  learning and its applications to wifi human sensing: A benchmark and a
  tutorial,'' \emph{arXiv preprint arXiv:2207.07859}, 2022.

\bibitem{yang2021deep}
J.~Yang, H.~Zou, L.~Xie, and C.~J. Spanos, ``Deep learning and unsupervised
  domain adaptation for wifi-based sensing,'' in \emph{Generalization With Deep
  Learning: For Improvement On Sensing Capability}, 2021, pp. 79--100.

\bibitem{zou2018deepsense}
H.~Zou, Y.~Zhou, J.~Yang, H.~Jiang, L.~Xie, and C.~J. Spanos, ``Deepsense:
  Device-free human activity recognition via autoencoder long-term recurrent
  convolutional network,'' in \emph{2018 IEEE International Conference on
  Communications (ICC)}.\hskip 1em plus 0.5em minus 0.4em\relax IEEE, 2018, pp.
  1--6.

\bibitem{wang2021multimodal}
D.~Wang, J.~Yang, W.~Cui, L.~Xie, and S.~Sun, ``Multimodal csi-based human
  activity recognition using gans,'' \emph{IEEE Internet of Things Journal},
  vol.~8, no.~24, pp. 17\,345--17\,355, 2021.

\bibitem{hassija2019survey}
V.~Hassija, V.~Chamola, V.~Saxena, D.~Jain, P.~Goyal, and B.~Sikdar, ``A survey
  on iot security: application areas, security threats, and solution
  architectures,'' \emph{IEEE Access}, vol.~7, pp. 82\,721--82\,743, 2019.

\bibitem{zheng2016improving}
S.~Zheng, Y.~Song, T.~Leung, and I.~Goodfellow, ``Improving the robustness of
  deep neural networks via stability training,'' in \emph{Proceedings of the
  ieee conference on computer vision and pattern recognition}, 2016, pp.
  4480--4488.

\bibitem{kannan2018adversarial}
H.~Kannan, A.~Kurakin, and I.~Goodfellow, ``Adversarial logit pairing,''
  \emph{arXiv preprint arXiv:1803.06373}, 2018.

\bibitem{hendrycks2019augmix}
D.~Hendrycks, N.~Mu, E.~D. Cubuk, B.~Zoph, J.~Gilmer, and B.~Lakshminarayanan,
  ``Augmix: A simple data processing method to improve robustness and
  uncertainty,'' in \emph{International Conference on Learning
  Representations}, 2019.

\bibitem{samangouei2018defense}
P.~Samangouei, M.~Kabkab, and R.~Chellappa, ``Defense-gan: Protecting
  classifiers against adversarial attacks using generative models,''
  \emph{arXiv preprint arXiv:1805.06605}, 2018.

\bibitem{meng2017magnet}
D.~Meng and H.~Chen, ``Magnet: a two-pronged defense against adversarial
  examples,'' in \emph{Proceedings of the 2017 ACM SIGSAC conference on
  computer and communications security}, 2017, pp. 135--147.

\bibitem{wang2020CMT}
D.~Wang, C.~Li, S.~Wen, S.~Nepal, and Y.~Xiang, ``Defending against adversarial
  attack towards deep neural networks via collaborative multi-task training,''
  \emph{IEEE Transactions on Dependable and Secure Computing}, pp. 1--1, 2020.

\bibitem{van2008visualizing}
L.~Van~der Maaten and G.~Hinton, ``Visualizing data using t-sne.''
  \emph{Journal of machine learning research}, vol.~9, no.~11, 2008.

\bibitem{wang2017device}
W.~Wang, A.~X. Liu, M.~Shahzad, K.~Ling, and S.~Lu, ``Device-free human
  activity recognition using commercial wifi devices,'' \emph{IEEE Journal on
  Selected Areas in Communications}, vol.~35, no.~5, pp. 1118--1131, 2017.

\bibitem{zou2018identification}
H.~Zou, Y.~Zhou, J.~Yang, W.~Gu, L.~Xie, and C.~Spanos, ``Wifi-based human
  identification via convex tensor shapelet learning,'' pp. 1711--1719, 2018.

\bibitem{zou2018joint}
H.~Zou, J.~Yang, Y.~Zhou, and C.~J. Spanos, ``Joint adversarial domain
  adaptation for resilient wifi-enabled device-free gesture recognition,'' in
  \emph{2018 17th IEEE International Conference on Machine Learning and
  Applications (ICMLA)}.\hskip 1em plus 0.5em minus 0.4em\relax IEEE, 2018, pp.
  202--207.

\bibitem{wang2022caution}
D.~Wang, J.~Yang, W.~Cui, L.~Xie, and S.~Sun, ``Caution: A robust wifi-based
  human authentication system via few-shot open-set gait recognition,''
  \emph{IEEE Internet of Things Journal}, 2022.

\bibitem{yang2022metafi}
J.~Yang, Y.~Zhou, H.~Huang, H.~Zou, and L.~Xie, ``Metafi: Device-free pose
  estimation via commodity wifi for metaverse avatar simulation,'' \emph{arXiv
  preprint arXiv:2208.10414}, 2022.

\bibitem{nguyen2018eyelight}
V.~Nguyen, M.~Ibrahim, S.~Rupavatharam, M.~Jawahar, M.~Gruteser, and R.~Howard,
  ``Eyelight: Light-and-shadow-based occupancy estimation and room activity
  recognition,'' in \emph{IEEE INFOCOM 2018-IEEE Conference on Computer
  Communications}.\hskip 1em plus 0.5em minus 0.4em\relax IEEE, 2018, pp.
  351--359.

\bibitem{zou2018robust}
H.~Zou, J.~Yang, Y.~Zhou, L.~Xie, and C.~J. Spanos, ``Robust wifi-enabled
  device-free gesture recognition via unsupervised adversarial domain
  adaptation,'' in \emph{2018 27th International Conference on Computer
  Communication and Networks (ICCCN)}.\hskip 1em plus 0.5em minus 0.4em\relax
  IEEE, 2018, pp. 1--8.

\bibitem{hussain2020review}
Z.~Hussain, Q.~Z. Sheng, and W.~E. Zhang, ``A review and categorization of
  techniques on device-free human activity recognition,'' \emph{Journal of
  Network and Computer Applications}, vol. 167, p. 102738, 2020.

\bibitem{shi2020environment}
Z.~Shi, J.~A. Zhang, Y.~D.~R. Xu, and Q.~Cheng, ``Environment-robust
  device-free human activity recognition with channel-state-information
  enhancement and one-shot learning,'' \emph{IEEE Transactions on Mobile
  Computing}, 2020.

\bibitem{yang2022efficientfi}
J.~Yang, X.~Chen, H.~Zou, D.~Wang, Q.~Xu, and L.~Xie, ``Efficientfi: Towards
  large-scale lightweight wifi sensing via csi compression,'' \emph{IEEE
  Internet of Things Journal}, 2022.

\bibitem{yang2022autofi}
J.~Yang, X.~Chen, H.~Zou, D.~Wang, and L.~Xie, ``Autofi: Towards automatic wifi
  human sensing via geometric self-supervised learning,'' \emph{arXiv preprint
  arXiv:2205.01629}, 2022.

\bibitem{yang2020mobileda}
J.~Yang, H.~Zou, S.~Cao, Z.~Chen, and L.~Xie, ``Mobileda: Toward edge domain
  adaptation,'' \emph{IEEE Internet of Things Journal}, vol.~7, no.~8, pp.
  6909--6918, 2020.

\bibitem{wang2016human}
H.~Wang, D.~Zhang, J.~Ma, Y.~Wang, Y.~Wang, D.~Wu, T.~Gu, and B.~Xie, ``Human
  respiration detection with commodity wifi devices: do user location and body
  orientation matter?'' in \emph{Proceedings of the 2016 ACM International
  Joint Conference on Pervasive and Ubiquitous Computing}, 2016, pp. 25--36.

\bibitem{zhang2021wi}
L.~Zhang, C.~Wang, and D.~Zhang, ``Wi-pigr: Path independent gait recognition
  with commodity wi-fi,'' \emph{IEEE Transactions on Mobile Computing}, 2021.

\bibitem{hu2022resfi}
J.~Hu, J.~Yang, J.-B. Ong, D.~Wang, and L.~Xie, ``Resfi: Wifi-enabled
  device-free respiration detection based on deep learning,'' in \emph{2022
  IEEE 17th International Conference on Control \& Automation (ICCA)}.\hskip
  1em plus 0.5em minus 0.4em\relax IEEE, 2022, pp. 510--515.

\bibitem{luo2020attackers}
Z.~Luo, S.~Zhao, Z.~Lu, J.~Xu, and Y.~Sagduyu, ``When attackers meet ai:
  Learning-empowered attacks in cooperative spectrum sensing,'' \emph{IEEE
  Transactions on Mobile Computing}, 2020.

\bibitem{sagduyu2019adversarial}
Y.~Sagduyu, Y.~Shi, and T.~Erpek, ``Adversarial deep learning for over-the-air
  spectrum poisoning attacks,'' \emph{IEEE Transactions on Mobile Computing},
  2019.

\bibitem{dong2019efficient}
Y.~Dong, H.~Su, B.~Wu, Z.~Li, W.~Liu, T.~Zhang, and J.~Zhu, ``Efficient
  decision-based black-box adversarial attacks on face recognition,'' in
  \emph{Proceedings of the IEEE/CVF Conference on Computer Vision and Pattern
  Recognition}, 2019, pp. 7714--7722.

\bibitem{li2020bert}
L.~Li, R.~Ma, Q.~Guo, X.~Xue, and X.~Qiu, ``Bert-attack: Adversarial attack
  against bert using bert,'' \emph{arXiv preprint arXiv:2004.09984}, 2020.

\bibitem{yang2020adversarial}
Z.~Yang, Y.~Zhao, and W.~Yan, ``Adversarial vulnerability in doppler-based
  human activity recognition,'' in \emph{2020 International Joint Conference on
  Neural Networks (IJCNN)}.\hskip 1em plus 0.5em minus 0.4em\relax IEEE, 2020,
  pp. 1--7.

\bibitem{huang2021wars}
P.~Huang, X.~Zhang, S.~Yu, and L.~Guo, ``Is-wars: Intelligent and stealthy
  adversarial attack to wi-fi-based human activity recognition systems,''
  \emph{IEEE Transactions on Dependable and Secure Computing}, 2021.

\bibitem{peng2020suppressing}
X.~Peng, K.~Wang, Z.~Zeng, Q.~Li, J.~Yang, and Y.~Qiao, ``Suppressing
  mislabeled data via grouping and self-attention,'' in \emph{European
  Conference on Computer Vision}, 2020.

\end{thebibliography}


%
\begin{IEEEbiography}[{\includegraphics[width=0.9in,clip,keepaspectratio]{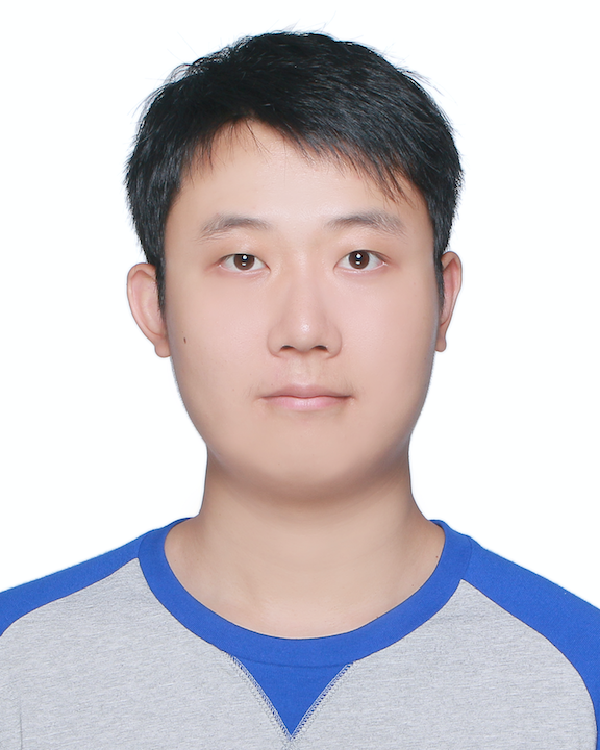}}]{Jianfei Yang} received the B.Eng. from the School of Data and Computer Science, Sun Yat-sen University in 2016, and the Ph.D. degree from Nanyang Technological University (NTU), Singapore in 2021. He received the best Ph.D. thesis award from NTU. He used to work as a senior research engineer at BEARS, the University of California, Berkeley. His research interests include deep learning, wireless sensing, and artificial intelligence of things. He won many International AI challenges in computer vision and interdisciplinary research fields. Currently, he is a Presidential Postdoctoral Research Fellow and an independent PI at NTU.
\end{IEEEbiography}

\begin{IEEEbiography}[{\includegraphics[width=1in,height=1.25in,clip,keepaspectratio]{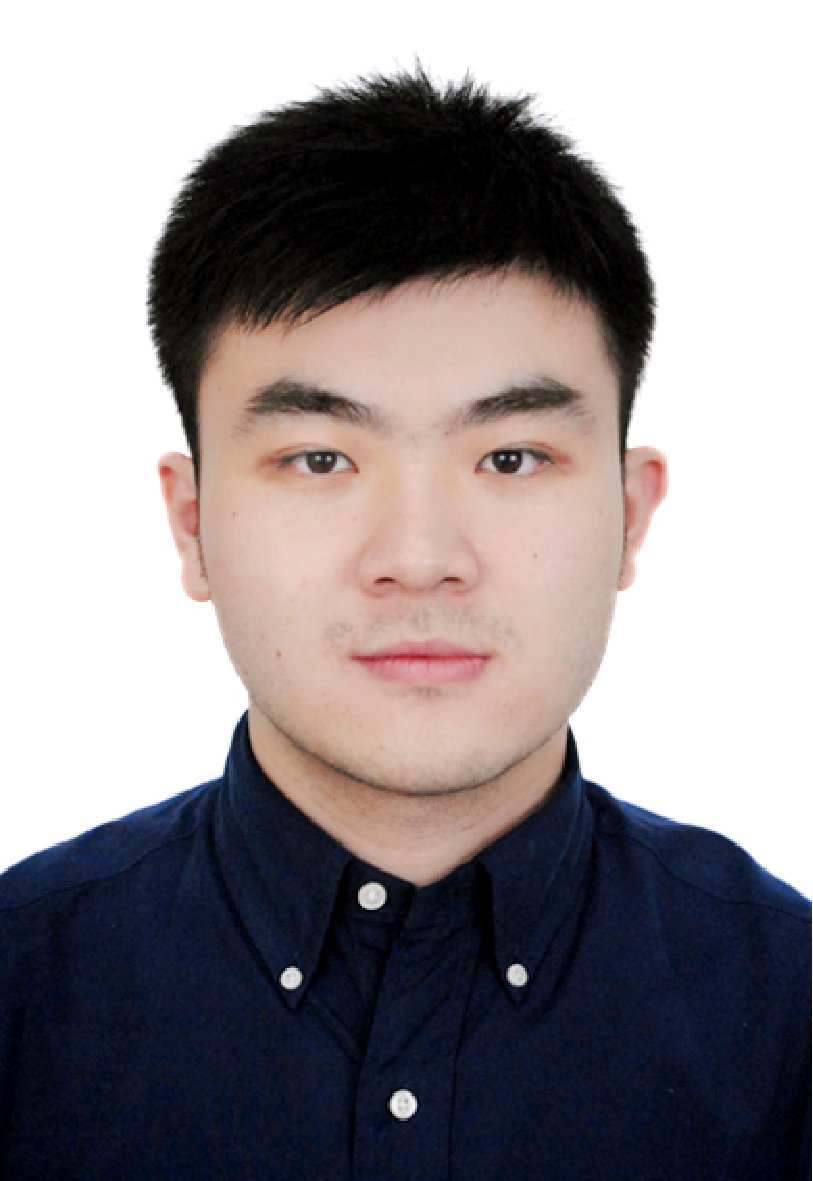}}]{Han Zou}
	received the B.Eng. (First Class Honors) and Ph.D. degrees in Electrical and Electronic Engineering from the Nanyang Technological University, Singapore, in 2012 and 2016, respectively. He is currently a Postdoctoral Scholar with the Department of Electrical Engineering and Computer Sciences at the University of California, Berkeley, CA, USA. His research interests include ubiquitous computing, statistical learning, signal processing and data analytics with applications in occupancy sensing, indoor localization, smart buildings and Internet of Things.	
\end{IEEEbiography}
\begin{IEEEbiography}[{\includegraphics[width=1in,height=1.25in,clip,keepaspectratio]{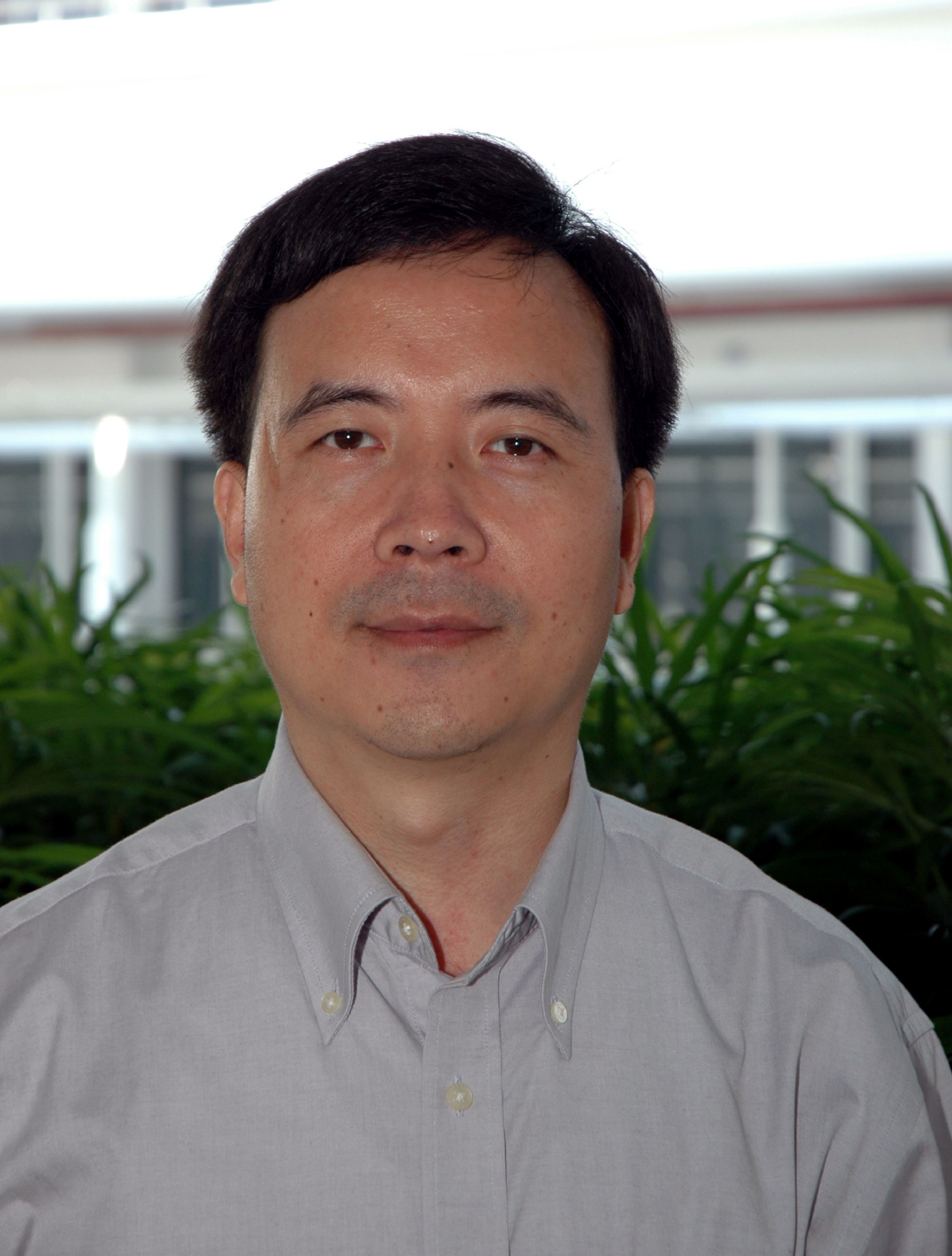}}]{Lihua Xie}
	received the B.E. and M.E. degrees in electrical engineering from Nanjing University of Science and Technology in 1983 and 1986, respectively, and the Ph.D. degree in electrical engineering from the University of Newcastle, Australia, in 1992. Since 1992, he has been with the School of Electrical and Electronic Engineering, Nanyang Technological University, Singapore, where he is currently a professor and served as the Head of Division of Control and Instrumentation from July 2011 to June 2014. He held teaching appointments in the Department of Automatic Control, Nanjing University of Science and Technology from 1986 to 1989 and Changjiang Visiting Professorship with South China University of  Technology from 2006 to 2011.
	
	Dr Xie's research interests include robust control and estimation, networked control systems, multi-agent control and unmanned systems. He has served as an editor of IET Book Series in Control and an Associate Editor of a number of journals including IEEE Transactions on Automatic Control, Automatica, IEEE Transactions on Control Systems Technology, and IEEE Transactions on Circuits and Systems-II. Dr Xie is a Fellow of Academy of Engineering Singapore, Fellow of IEEE, Fellow of IFAC, and Fellow of Chinese Automation Association.
\end{IEEEbiography}

\end{document}